\documentclass[12pt,preprint]{aastex}
\begin{document}
\title{Revisiting the Galaxy Shape and  Spin Alignments with the Large-Scale Tidal Field: An Effective Practical Model}
\author{Jounghun Lee}
\affil{Astronomy Program, Department of Physics and Astronomy, Seoul National University, 
Seoul 08826, Republic of Korea} 
\email{jounghun@astro.snu.ac.kr}
%%%%%%%%%%%%%%%%%%%%%%%%%%%%%%%%%%%%%%%%%%%%%%%%%%%%%%%%%%%%%%%%
\begin{abstract}
An effective practical model with two characteristic parameters is presented to describe both of the tidally induced shape and spin 
alignments of the galactic halos with the large-scale tidal fields.  We test this model against the numerical results obtained from the 
Small MultiDark Planck simulation on the galactic mass scale of $0.5\le M/(10^{11}\,h^{-1}\,M_{\odot})\le 50$ at redshift $z=0$. 
Determining empirically the parameters from the numerical data, we demonstrate how successfully our model describes simultaneously 
and consistently the amplitudes and behaviors of the probability density functions of three coordinates of the shape and spin vectors in 
the principal frame of the large scale tidal field. Dividing the samples of the galactic halos into multiple subsamples in four different mass
ranges and four different types of the cosmic web, and also varying the smoothing scale of the tidal field from 
$5\,h^{-1}$Mpc to $10,\ 20,\ 30\,h^{-1}$Mpc, we perform repeatedly the numerical tests with each subsample at each scale. 
Our model is found to match well the numerical results for all of the cases of the mass range, smoothing scale and web type 
and to properly capture the scale and web dependence of the spin flip phenomenon. 
\end{abstract}
%%%%%%%%%%%%%%%%%%%%%%%%%%%%%%%%%%%%%%%%%%%%%%%%%%%%%%%%%%%%%%%%

\section{Introduction}\label{sec:intro}

The physical properties of the observed galaxies in the universe is a reservoir of information on the conditions under which 
they formed, the evolutionary processes which they went through, and the interactions in which they are involved.  
Although the local conditions and processes at the galactic scales must have had the most dominant impact on the galaxies, 
the non-local effects beyond the galactic scales are also believed to have contributed  partly to their physical properties 
\citep[e.g.,][]{PS17}.  Subdominant as its contribution is,  the non-local effects on the galaxies are worth investigating, 
since it may contain valuable independent information on the galaxy formation and the background cosmology as well. 

The non-local effects on the galaxies are manifested by the correlations between the galaxy properties and the large-scale 
environments. Among various properties of the galaxies that have been found correlated with the large-scale environments, 
the shape and spin alignments of the galaxies with the large-scale structures (collectively called the galaxy intrinsic alignments) 
have lately drawn considerable attentions, inspiring vigorous extensive studies 
\citep[see][for recent reviews]{review1,review2,review3}.  
It is partially because the galaxy intrinsic alignments, if present and significant, could become another systematics in the 
measurements of the extrinsic counterparts caused by the weak gravitational lensing \citep[see][and references therein]{TI15}. 
The other important motivation for the recent flurry of research on this topic is that the origin of the galaxy intrinsic alignments 
is amenable to the first order perturbation theory and thus a rather fundamental approach to this topic is feasible 
\citep[e.g.,][]{hea-etal00,LP00,cat-etal01,cri-etal01,LP01,por-etal02,HZ08,bla-etal11,bla-etal15,tug-etal18}. 

In the first order Lagrangian perturbation theory \citep{zel70,buc92}, the minor (major) eigenvectors of the inertia momentum 
tensors of the proto-galactic regions are perfectly aligned with the major (minor) eigenvectors of the local tidal tensors around the 
regions. Several $N$-body simulations have indeed detected the existence of strong correlations between the inertia momentum and 
local tidal tensors at the proto-galactic sites \citep{LP00,por-etal02,lee-etal09}. Since the tidal fields smoothed on different 
scales are cross correlated, the eigenvectors of the inertia momentum tensors of the proto-galactic regions are expected to be 
aligned with those of the large-scale tidal fields.  The major eigenvectors of the inertia momentum tensors of the proto-galaxies 
correspond to the most elongated axes of their shapes, while the minor eigenvectors of the large-scale tidal tensors correspond 
to the directions along which the surrounding matter become minimally compressed. Henceforth, this expectation based on the first 
order Lagrangian perturbation theory basically translates into the possible alignments between the galaxy shapes and the 
most elongated axes of the large-scale structures such as the axes of the filaments, the signals of which have been detected by 
several numerical and observational studies \citep[e.g.,][and references therein]{alt-etal06,hah-etal07b,zha-etal09,zha-etal13,che-etal16}.

In the linear tidal torque (LTT) theory that \citet{dor70} formulated by combining the first order Lagrangian perturbation theory with 
the Zel'dovich approximation \citep{zel70}, the anisotropic tidal field of the surrounding matter distribution originates the spin 
angular momentum of a proto-galaxy provided that its shape departs from a spherical symmetry. The generic and unique prediction 
of this LTT theory is the inclinations of the spin vectors of the proto-galaxies toward the intermediate eigenvectors of the large scale 
tidal field \citep{LP00}, which has also garnered several numerical and observational supports 
\citep[e.g.,][]{nav-etal04,tru-etal06,hah-etal07a,LE07,wan-etal11,zha-etal15,che-etal16}.

The recently available large high-resolution $N$-body simulations that covered a broad mass range, however, limited the validity of 
the LTT prediction  to the mass scale of $M\ge M_{\rm t}\sim 10^{12}\,h^{-1}\,M_{\odot}$, showing that on the mass scale below 
$M_{t}$ the spin vectors of dark matter halos at $z=0$ are aligned not with the intermediate but rather with the minor eigenvectors of 
the large scale tidal field, similar to the axes of the halo shapes 
\citep{ara-etal07,hah-etal07b,paz-etal08,zha-etal09,cod-etal12,lib-etal13,tro-etal13,dub-etal14,vee-etal18}. 
This difference in the spin alignment tendency between the low and high mass scales were found most conspicuous 
in the filament environments: the spin axes of the galactic halos with masses lower (higher) than $M_{\rm t}$ measured at $z=0$ 
tend to be parallel (perpendicular) to the elongated axes of their host filaments, in contradiction with the LTT prediction. 
The transition of the spin alignment tendency at $M_{\rm t}$ is often called "spin flip" phenomenon \citep{cod-etal12} and the 
break-down of the LTT prediction below $M_{\rm t}$ has also been witnessed in recent observations 
\citep{tem-etal13,TL13,hir-etal17,che-etal18}. 

The detection of this spin-flip phenomenon puzzled the community and urged it to find a proper answer to the critical question of 
what the origin of this phenomenon is. What has so far been suggested as a possible origin includes the major merging events, 
mass dependence of the merging and accretion processes, assembly bias, vorticity generation inside filaments, web-dependence 
of the galaxy formation epochs, nonlinear tidal interactions, geometrical properties of the host filaments and etc 
\citep{BF12,LP12,cod-etal12,lib-etal13,wel-etal14,cod-etal15,lai-etal15,BF16,WK17,vee-etal18}. 
Although these previously suggested factors were believed to play some roles for the occurrence of the spin-flip phenomenon, 
none of them are fully satisfactory in explaining all aspects of the spin-flip phenomenon including the dependence of the transition 
mass scale $M_{\rm t}$ on the types of the cosmic web, redshifts, and scales of the filaments. 

The occurrence of the spin-flip phenomenon basically implies that for the case of the galaxies with masses $M\le M_{t}$, the tendency of 
the spin alignments with the large scale tidal field becomes similar to that of the shape alignments. Thus, it is suspected that whatever 
caused the spin-flip phenomenon, it should be linked to the shape alignments with the large scale tidal field.  To address these remaining 
issues, what is highly desired is an effective model that can describe consistently and simultaneously both of the galaxy shape and spin 
alignments. Here, we attempt to construct such a model by modifying the original LTT theory and to explore if the 
shapes of the galaxies also show any transition of the alignment tendency like the spin counterparts

The organization of this Paper is as follows. A refined analytic model for the galaxy shape alignments is presented in Section 
\ref{sec:shape_model} and tested against the numerical results in Section \ref{sec:shape_test}. An effective model for the tidally 
induced spin alignments  is presented in Section \ref{sec:spin_model} and tested against the numerical results in Section 
\ref{sec:spin_test}. A discussion over the possible application of this model as well as a summary of the results is presented in 
Section \ref{sec:con}. Throughout this Paper, we will assume a Planck universe whose total energy density is dominantly contributed 
by the cosmological constant ($\Lambda$) and the cold dark matter (CDM) \citep{planck13}. 

\section{Tidally Induced Shape Alignments}\label{sec:shape}

\subsection{An Analytic Model}\label{sec:shape_model}

Suppose a galactic halo located in a region where a tidal tensor ${\bf T}$ has its major, intermediate and minor 
eigenvectors ($\hat{\bf u}_{1},\ \hat{\bf u}_{2}$ and $\hat{\bf u}_{3}$, respectively), corresponding to the largest, second to the largest, 
smallest eigenvalues ($\lambda_{1},\ \lambda_{2}$ and $\lambda_{3}$, respectively).  
The tidal tensor ${\bf T}$ depends on the smoothing scale, $R_{f}$, as 
${\bf T}({\bf x})\propto \partial_{i}\partial_{j}\int d{\bf x}^{\prime}\,\Phi({\bf x^{\prime}})W(\vert{\bf x}-{\bf x}^{\prime}\vert;R_{f})$, 
where $\Phi({\bf x})$ is the perturbation potential field and $W(\vert{\bf x}-{\bf x}^{\prime}\vert;R_{f})$ is a window function with 
a filtering radius $R_{f}$. In the current analysis, we adopt a Gaussian window function. 

As mentioned in Section \ref{sec:intro}, the first order Lagrangian perturbation theory \citep{zel70,buc92} predicts a strong anti-correlation 
between the principal axes of the inertia momentum tensor of a galactic halo and the local tidal tensor in the Lagrangian regime. 
According to this theory, the correlation between the two tensors is strongest if the two tensors are defined on the same scale (i.e., the 
virial radius of the halo, $R_{g}$), becomes weaker if $R_{f}$ is larger than $R_{g}$. If the shape of this galactic halo can be 
approximated by an ellipsoid, then the direction of the coordinate vector of the largest shape ellipsoid, ${\bf e}=({e}_{1}, {e}_{2}, {e}_{3})$ 
(i.e., the major principal axis of the inertia momentum tensor) is expected to be aligned with $\hat{\bf u}_{3}$ (i.e., the minor 
principal axis of the local tidal tensor) along which the surrounding matter is least compressed, provided that 
$\lambda_{1}\ne\lambda_{2}\ne\lambda_{3}$.

This alignment tendency can be statistically quantified by the conditional joint probability density function of three coordinates of the 
largest shape ellipsoid axis, $p({e}_{1}, {e}_{2}, {e}_{3}\vert \hat{\bf T})$, where 
$\hat{\bf T}$ is a unit {\it traceless} tidal tensor defined as $\hat{\bf T}\equiv ({\bf T}-{\rm Tr}({\bf T})/3)/\vert {\bf T}-{\rm Tr}({\bf T})/3\vert$ 
with ${\rm Tr}({\bf T})$ denoting the trace of ${\bf T}$. 
As \citet{LP01} and \citet{lee04} did, we assume here that $p({e}_{1}, {e}_{2}, {e}_{3}\vert \hat{\bf T})$ follows a multivariate Gaussian 
distribution of 
%%%%%%%%%%%%%%%%%%%%%%%%%%%%%%%%%%%%%%%%%%%%%%%%%%%%%%%%%%%%%%%
\begin{equation}
\label{eqn:jpro_axis}
p({e}_{1},\ {e}_{2},\ {e}_{3}\vert \hat{\bf T}) = \frac{1}{[(2\pi)^3 {\rm det}({\bf \Sigma})]^{1/2}}
\exp\left[-\frac{1}{2}\left({\bf e}\cdot{\bf \Sigma}^{-1}\cdot{\bf e}\right)\right]\, , 
\end{equation}
%%%%%%%%%%%%%%%%%%%%%%%%%%%%%%%%%%%%%%%%%%%%%%%%%%%%%%%%%%%%%%%
where the components of the covariance matrix, ${\bf \Sigma}=({\Sigma}_{ij})$, are defined as the conditional 
ensemble averages, ${\Sigma}_{ij} \equiv \langle{e}_{i}{e}_{j}\vert\hat{\bf T}\rangle$. 
Here, we suggest the following practical formula for $\langle{e}_{i}{e}_{j}\vert\hat{\bf T}\rangle$:  
%%%%%%%%%%%%%%%%%%%%%%%%%%%%%%%%%%%%%%%%%%%%%%%%%%%%%%%%%%%%%%%
\begin{equation}
\label{eqn:eiej}
\langle{e}_{i}{e}_{j}\vert\hat{\bf T}\rangle = \frac{1+d_{t}}{3}\delta_{ij} - d_{t}\hat{T}_{ij}\, ,
\end{equation}
%%%%%%%%%%%%%%%%%%%%%%%%%%%%%%%%%%%%%%%%%%%%%%%%%%%%%%%%%%%%%%%
where $d_{t}$ is the shape correlation parameter that measures the alignment strength between ${\bf e}$ and $\hat{\bf u}_{3}$. 
Note that this formula describes a linear dependence of the covariance, $\langle{e}_{i}{e}_{j}\vert\hat{\bf T}\rangle$, on ${\bf T}$ 
\citep{cat-etal01,LP08,HZ08}, unlike a spin vector whose covariance has a quadratic dependence on ${\bf T}$ 
\citep{whi84,LP00,LP01}.

Focusing only on the direction of ${\bf e}$, we marginalize $p({e}_{1}, {e}_{2}, {e}_{3}\vert {\bf T})$ over 
$e\equiv\vert{\bf e}\vert$ to have 
%%%%%%%%%%%%%%%%%%%%%%%%%%%%%%%%%%%%%%%%%%%%%%%%%%%%%%%%%%%%%%%
\begin{eqnarray}
\label{eqn:hjpro_axis}
p(\hat{\bf e}\vert\hat{\bf T})&=&\int\,p({\bf e}\vert\hat{\bf T}) e^{2}de \nonumber \\
&=&\frac{1}{4\pi{\rm det}({\bf\Sigma})^{1/2}}\left(\hat{\bf e}\cdot{\bf \Sigma}^{-1}\cdot\hat{\bf e}\right)^{-3/2}\, , 
%&=&\frac{1}{4\pi}\left[1+\frac{3d_{t}}{2}\left(1-3\hat{T}_{ij}\hat{e}_{i}\hat{e}_{j}\right)\right]\, ,
\end{eqnarray}
%%%%%%%%%%%%%%%%%%%%%%%%%%%%%%%%%%%%%%%%%%%%%%%%%%%%%%%%%%%%%%%
where $\hat{\bf e}\equiv {\bf e}/e$ denotes the unit vector in the direction of the largest shape ellipsoid axis. 
While $\hat{\bf T}$ shares the same orthonormal eigenvectors with  ${\bf T}$, its eigenvalues, 
$\hat{\lambda}_{1},\ \hat{\lambda}_{2},\ \hat{\lambda}_{3}$, are subject to two additional constraints  of 
$\sum_{i=1}^{3}\hat{\lambda}_{i}=0$ and $\sum_{i=1}^{3}\hat{\lambda}^{3}_{i}=1$ \citep{LP01}.

Putting Equation (\ref{eqn:eiej}) into Equation (\ref{eqn:hjpro_axis}) leads to the following analytic expression 
%%%%%%%%%%%%%%%%%%%%%%%%%%%%%%%%%%%%%%%%%%%%%%%%%%%%%%%%%%%%
\begin{eqnarray}
\label{eqn:hjpro_axis_dt}
p(\hat{\bf e}\vert\hat{\bf T}) 
&=& = \frac{1}{2\pi}\left[\prod_{n=1}^{3}\left(1+d_{t}-3d_{t}\hat{\lambda}_{n}\right)\right]^{-\frac{1}{2}}
\left(\sum_{l=1}^{3}\frac{\vert\hat{\bf u}_{l}\cdot\hat{\bf e}\vert}{1+d_{t}-3d_{t}\hat{\lambda}_{l}} \right)^{-\frac{3}{2}}\, ,
\end{eqnarray}
%%%%%%%%%%%%%%%%%%%%%%%%%%%%%%%%%%%%%%%%%%%%%%%%%%%%%%%%%%%
since $\hat{T}_{ij}=\hat{\lambda}_{i}\delta_{ij}$ in the principal axis frame of $\hat{\bf T}$. 
Now, the conditional probability density function, $p(\vert\hat{\bf u}_{i}\cdot\hat{\bf e}\vert)$, for $i\in \{1,2,3\}$, 
can be obtained as
%%%%%%%%%%%%%%%%%%%%%%%%%%%%%%%%%%%%%%%%%%%%%%%%%%%%%%%%%%%%
\begin{equation}
\label{eqn:hpro_axis_dt}
p(\vert\hat{\bf u}_{i}\cdot\hat{\bf e}\vert)=\int_{0}^{2\pi}p(\hat{\bf e}\vert\hat{\bf T})\,d\phi_{jk}\, ,
\end{equation}
%%%%%%%%%%%%%%%%%%%%%%%%%%%%%%%%%%%%%%%%%%%%%%%%%%%%%%%%%%%
where $\phi_{jk}$ is the azimuthal angle of $\hat{\bf e}$ in the plane spanned by $\hat{\bf u}_{j}$ and $\hat{\bf u}_{k}$ 
perpendicular to $\hat{\bf u}_{i}$.

Equation (\ref{eqn:hjpro_axis_dt}) indicates that the completion of this analytic model requires us to determine the value of $d_{t}$. 
If $\hat{\bf e}$ is perfectly aligned with $\hat{\bf u}_{3}$, then $d_{t}$ would be unity. Whereas, the zero value of $d_{t}$ would 
correspond to the case that $\hat{\bf e}$ is completely random having no correlation with $\hat{\bf u}_{3}$.  
As done in \citet{LP01}, for the determination of $d_{t}$, we first evaluate the conditional ensemble average, 
$\langle\hat{e}_{i}\hat{e}_{j}\vert \hat{\bf T}\rangle$, under the assumption of $d_{t}\ll 1$
%%%%%%%%%%%%%%%%%%%%%%%%%%%%%%%%%%%%%%%%%%%%%%%%%%%%%%%%%%%%%%%
\begin{eqnarray}
\label{eqn:hei_def}
\langle\hat{e}_{i}\hat{e}_{j}\vert \hat{\bf T}\rangle &=& \int \hat{e}_{i}\hat{e}_{j} p(\hat{\bf e}\vert\hat{\bf T}) d\hat{\bf e}\, ,\\
\label{eqn:hei}
&\approx& \left[\left(\frac{1}{3}+\frac{3}{5}d_{t}\right) - \frac{3}{5}d_{t}\hat{\lambda}_{i}\right]\delta_{ij}\, , 
\end{eqnarray}
%%%%%%%%%%%%%%%%%%%%%%%%%%%%%%%%%%%%%%%%%%%%%%%%%%%%%%%%%%%%%%%
Note that the off-diagonal elements vanish in Equation (\ref{eqn:hei}) since $\hat{T}_{ij}=\hat{\lambda}_{i}\delta_{ij}$ in its principal 
frame. Multiplying Equation (\ref{eqn:hei}) by $\hat{\lambda}_{i}$ and summing over the three components, we finally derive 
a simple analytic formula for $d_{t}$: 
%%%%%%%%%%%%%%%%%%%%%%%%%%%%%%%%%%%%%%%%%%%%%%%%%%%%%%%%%%%%%%%
\begin{equation}
\label{eqn:dt_sol}
d_{t} = -\frac{5}{3}\sum_{i=1}^{3}\hat{\lambda}_{i}\langle\hat{e}^{2}_{i}\vert\hat{\bf T}\rangle\, .
\end{equation}
%%%%%%%%%%%%%%%%%%%%%%%%%%%%%%%%%%%%%%%%%%%%%%%%%%%%%%%%%%%%%%%
The constraints of $\sum_{i=1}^{3}\hat{\lambda}_{i}=0$ and $\sum_{i=1}^{3}\hat{\lambda}^{2}_{i}=1$ are used to derive the 
above formula. Equation (\ref{eqn:dt_sol}) implies that 
once the values of $\hat{\lambda}_{i}$ and $\langle\hat{e}^{2}_{i}\vert\hat{\bf T}\rangle$ are measured, the shape correlation parameter, 
$d_{t}$, can be empirically determined.

It is worth recalling that the shape correlation parameter, $d_{t}$, depends on the smoothing scale, $R_{f}$.  It is expected to have the 
highest value when $R_{f}=R_{g}$, as mentioned in the above. 
In the Eulerian regime, however, the approximation of $p({\bf e}\vert{\bf T})$ as a multivariate Gaussian distribution and ${\bf T}$ 
as a Gaussian random field used for Equation (\ref{eqn:hjpro_axis}) are not valid on the scale of $R_{g}$ due to the nonlinear evolution 
of $\hat{\bf T}$ on the galactic scale. Thus, we consider the scales $R_{f}$ much larger than $R_{g}$ where these approximations still hold 
true. Since $\hat{\bf T}$ on two different scales of $R_{g}$ and $R_{f}$ are cross-correlated, it is expected that $\hat{\bf e}$ is still 
correlated with $\hat{\bf T}$  smoothed on the scale of $R_{f}\gg R_{g}$. The larger the difference between $R_{f}$ and $R_{g}$ is, 
the lower the value of $d_{t}$ is. 

\subsection{Numerical Tests}\label{sec:shape_test}

Our numerical analysis is based on the data set from the Small MultiDark Planck 
simulation\footnote{doi:10.17876/cosmosim/smdpl/}(SMDPL), a DM only $N$-body simulation conducted in a periodic box with a 
side length of $400\,h^{-1}$Mpc \citep{smdpl} as a part of the MultiDark simulation project \citep{multidark} 
for a Planck universe \citep{planck13}. 
The SMDPL tracks down the gravitational evolution of $3840^{3}$ DM particles each of which has individual 
mass of $9.63\times 10^{7}\,h^{-1}\,M_{\odot}$, starting from $z=120$ down to $z=0$ \citep{smdpl}.  The virialized DM halos 
were identified via the Rockstar halo-finding algorithm \citep{rockstar} from the spatial distributions of the DM particles at various 
snapshots of the SMDPL.  

Through the CosmoSim database that stores all the experimental results from the MultiDark simulations, 
we first extract the Rockstar catalog, which provides information on a diverse set of the physical properties of the DM halos. 
For the current analysis, we use such information as the parent id (pId), comoving position vector (${\bf r}$), 
spatial grid index, virial mass ($M$), coordinate vector of the largest shape ellipsoid axis (${\bf e}$) of each Rockstar halo. 
The integer value of pId is used to exclude the subhalos from our analysis. For the case of a distinct halo that is not a subhalo 
hosted by any other larger halo, the parent id has the value of pId=-1. The coordinate vector, ${\bf e}$, is a measure of the most 
elongated axis of an ellipsoid to which the shape of a given Rockstar halo was fitted. 
From here on, the unit coordinate vector of the largest shape ellipsoid axis, $\hat{\bf e}\equiv {\bf e}/e$, will be called 
{\it a shape vector}. 

We make a sample of the {\it distinct galactic halos} by selecting only those from the Rockstar catalog 
which meet two conditions of pId$=-1$ and $0.5\le M/(10^{11}\,h^{-1}\,M_{\odot})<50$. 
Then, we divide this sample into four subsamples which cover four different mass ranges: 
$0.5\le M/(10^{11}\,h^{-1}\,M_{\odot})<1$ (lowest-mass galactic halos), 
$1\le M/(10^{11}\,h^{-1}\,M_{\odot})<5$ (low-mass galactic halos), 
$5\le M/(10^{11}\,h^{-1}\,M_{\odot})<10$ (medium-mass galactic halos) and 
$10\le M/(10^{11}\,h^{-1}\,M_{\odot})<50$ (high-mass galactic halos), respectively. 
We exclude the subhalos from the analysis, since the effect of the large-scale tidal field on the subhalos are likely to be 
negligible compared with that of the internal nonlinear tidal fields inside the host halos  
Those halos with $M<0.5\times 10^{11}\,h^{-1}\,M_{\odot}$ are excluded on the ground that the measurements of the shape and spin 
vectors of those halos are likely to be contaminated by the shot noise due to the small number of the component DM particles 
\citep{bet-etal07}. The group and cluster size halos with $M\ge 5\times 10^{12}\,h^{-1}\,M_{\odot}$ are also excluded since the 
measurements of their shapes and spins should be severely affected by their dynamical states, internal structures and recent 
merging events. 

We also retrieve the cloud-in-cell density field, $\rho({\bf r})$, defined on the $512^{3}$ grids at $z=0$ via the CosmoSim database.  
Then, we calculate the dimensionless density contrast field, 
$\delta({\bf r})=\left[\rho({\bf r})-\langle\rho\rangle\right]/\langle\rho\rangle$, where the 
ensemble average, $\langle\rho\rangle$, is taken over all the grids. 
With the help of the numerical recipe code that performs the Fast Fourier Transformation (FFT) \citep{recipe}, 
we compute the Fourier amplitude of the density contrast field,  $\tilde{\delta}({\bf k})$, where ${\bf k}=k(\hat{k}_{i})$ is the 
wave vector in the Fourier space.  
The inverse FFT of $\tilde{T}_{ij} = \hat{k}_{i}\hat{k}_{j}\tilde{\delta}({\bf k})\exp\left(-k^{2}R^{2}_{f}/2\right)$ for $i,j\in \{1,2,3\}$ 
leads us to have the tidal field, ${\bf T}({\bf r})$, smoothed on the scale of $R_{f}$.

For each subsample, we take the following steps. 
First, at the grid where each halo is placed, we perform a similarity transformation of ${\bf T}({\bf r})$, 
to find its eigenvectors $\{\hat{\bf u}_{i}\}_{i=1}^{3}$ as well as the eigenvalues $\{\lambda_{i}\}_{i=1}^{3}$.  
Second, we calculate, $\{\vert\hat{\bf u}_{i}\cdot\hat{\bf e}\vert\}_{i=1}^{3}$, whose values lie in the range of $[0,1]$. 
Breaking this unit interval $[0,1]$ into seven bins with equal length of $\Delta=1/7$, we count the number of the galactic halos, 
$n_{h,i}$, whose values of $\vert\hat{\bf u}_{i}\cdot\hat{\bf e}\vert$ fall in each bin for each $i\in \{1,2,3\}$.  
Third, the probability densities of $\vert\hat{\bf u}_{i}\cdot\hat{\bf e}\vert$ at each bin are determined as 
$p(\vert\hat{\bf u}_{i}\cdot\hat{\bf e}\vert)=n_{h,i}/(N_{t}\,\Delta)$ where $N_{t}$ is the total number of the galactic halos 
contained in each subsample.  

Figure \ref{fig:pro_axis} plots $p(\vert\hat{\bf u}_{3}\cdot\hat{\bf e}\vert)$ (left panel), $p(\vert\hat{\bf u}_{2}\cdot\hat{\bf e}\vert)$ 
(middle panels) and $p(\vert\hat{\bf u}_{1}\cdot\hat{\bf e}\vert)$ (right panels) as filled circular dots for the cases of the 
lowest-mass (top panels), low-mass (second from the top panels), medium mass (second from the bottom panels), and high-mass 
(bottom panels) galactic halos.  
To obtain these results, we smooth $\hat{\bf T}({\bf r})$ on the scale of $R_{f}=5\,h^{-1}$Mpc. 
As can be seen, for all four subsamples, the shape vector, $\hat{\bf e}$, shows a strong inclination 
(anti-inclination) toward the minor (major) eigenvector, $\hat{\bf u}_{3}$ ($\hat{\bf u}_{1}$), 
while it shows no alignment with the intermediate eigenvector, $\hat{\bf u}_{2}$. 
Note also that the higher-mass galactic halos exhibit stronger alignment (anti-alignment) tendency between 
$\hat{\bf e}$ and $\hat{\bf u}_{3}$ ($\hat{\bf e}$ and $\hat{\bf u}_{1}$), which are consistent with the previously 
reported numerical and observational results 
\citep[e.g.,][and references therein]{hah-etal07b,zha-etal09,joa-etal13,zha-etal13,che-etal16,hil-etal17,xia-etal17,pir-etal18}.

To compare the analytic model presented in Section \ref{sec:shape_model} against these numerical results, we first calculate the 
mean values of $\hat{\lambda}_{i}$ and $d_{t}$ averaged over the galactic halos contained in each subsample as 
%%%%%%%%%%%%%%%%%%%%%%%%%%%%%%%%%%%%%%%%%%%%%%%%%%%%%%%%%%%%%%%
\begin{eqnarray}
\label{eqn:mhl}
\langle\hat{\lambda}_{i}\rangle &=& \frac{1}{N_{t}}\sum_{\alpha=1}^{N_{t}}\hat{\lambda}_{\alpha,i}=
\frac{1}{N_{t}}\sum_{\alpha=1}^{N_{t}}\, ,
\left(\frac{\tilde{\lambda}_{\alpha,i}}{\sqrt{\sum_{j=1}^{3}\tilde{\lambda}^{2}_{\alpha,j}}}\right)\, ,  \quad i\in \{1,2,3\}\, , \\
\label{eqn:tl}
\tilde{\lambda}_{\alpha,i} &=& \lambda_{\alpha,i} - \frac{1}{3}\sum_{j=1}^{3}\lambda_{\alpha,j}\, , \quad i\in \{1,2,3\}\, , \\
\label{eqn:mdt}
\langle d_{t}\rangle &=& \frac{1}{N_{t}}
\sum_{\alpha=1}^{N_{t}}\left[-\frac{5}{3}\sum_{i=1}^{3}\hat{\lambda}_{\alpha,i}
\vert\hat{\bf u}_{\alpha,i}\cdot\hat{\bf e}_{\alpha}\vert^{2}\right]\, ,
\end{eqnarray}
%%%%%%%%%%%%%%%%%%%%%%%%%%%%%%%%%%%%%%%%%%%%%%%%%%%%%%%%%%%%%%%
where $\{{\lambda}_{\alpha,i}\}_{i=1}^{3}$ denotes a set of the three eigenvalues of ${\bf T}$ at the grid where the $\alpha$th DM halo 
of a given subsample is located, and $\hat{\bf e}_{\alpha}$ is the shape vector of the $\alpha$th DM halo. 
Note that $\langle\hat{e}^{2}_{i}\vert\hat{\bf T}\rangle$ in Equation (\ref{eqn:dt_sol}) is approximated by 
$\vert\hat{\bf u}_{\alpha,i}\cdot\hat{\bf e}_{\alpha}\vert^{2}$ in Equation (\ref{eqn:mdt}) since the measured 
values in numerical realizations are believed to be close to the expectation values in theory. 

Substituting these mean values of $\langle\hat{\lambda}_{i}\rangle$ and $\langle d_{t}\rangle$ for $\hat{\lambda}_{i}$ and 
$d_{t}$ respectively in Equations (\ref{eqn:hjpro_axis_dt})-(\ref{eqn:hpro_axis_dt}), 
we evaluate the analytical model and plot them as red solid lines in Figure \ref{fig:pro_axis}. 
As can be seen, for all of the four cases of the halo mass ranges, the analytic model with the empirically determined parameter 
$d_{t}$ describes very well not only the alignments of $\hat{\bf e}$ with $\hat{\bf u}_{3}$ but also simultaneously its anti-alignment with 
$\hat{\bf u}_{1}$ and no correlation with $\hat{\bf u}_{2}$ as well, even though no fitting process is involved. 
Figure \ref{fig:dt_m} plots $\langle d_{t}\rangle$ for the four different cases of the mass ranges, showing quantitatively how the 
strength of the shape alignments increases with the  increment of $M$. 

Smoothing $\hat{\bf T}$ on three larger scales, $R_{f}=10,\ 20$ and $30\,h^{-1}$Mpc, we repeat the whole 
calculations, the results of which are shown in Figure \ref{fig:pro_axis_filter} for the case of the high-mass galactic halos.  
The analytic model with the empirically determined parameter $d_{t}$ agrees quite well with the numerical results 
for all of the three cases of $R_{f}$. Figure \ref{fig:dt_f} shows quantitatively how the increment of $R_{f}$ weakens the shape 
alignments. Although the alignment tendency becomes weaker as $R_{f}$ increases, the shape vector, $\hat{\bf e}$, still shows 
significant alignment (anti-alignment) with $\hat{\bf u}_{3}$ ($\hat{\bf u}_{1}$) even for the case of $R_{f}=30\,h^{-1}$Mpc, 
which is consistent with the findings of the previous works \citep[e.g.,][]{xia-etal17}. 

True as it is that our analytic model shows good quantitive agreements with the numerical results for the case of the high-mass 
galactic halos, it is not perfect. Some discrepancies are found in the behaviors of $p(\vert\hat{\bf u}_{3}\cdot\hat{\bf e}\vert)$ and 
$p(\vert\hat{\bf u}_{1}\cdot\hat{\bf e}\vert)$ between the analytic model and the numerical results, as can be seen in the bottom 
panel of Figure \ref{fig:pro_axis}.  The former describes a slightly milder increase of 
$p(\vert\hat{\bf u}_{3}\cdot\hat{\bf e}\vert)$ with $\vert\hat{\bf u}_{3}\cdot\hat{\bf e}\vert$ and a slightly milder decrease of 
$\vert\hat{\bf u}_{1}\cdot\hat{\bf e}\vert$ with $\vert\hat{\bf u}_{i}\cdot\hat{\bf e}\vert$ than the latter
especially for the case of the high-mass galactic halos. 
However, Figure \ref{fig:pro_axis_filter} shows that the increment of $R_{f}$ improves the agreements between the analytic 
model and the numerical results, which in turn implies that the discrepancies may be caused by the uncertainties associated with the 
approximations of $p({\bf e}\vert{\bf T})$ as a multivariate Gaussian distribution and ${\bf T}$ as a Gaussian random field made 
to derive the analytic model.  The larger the scales are, the more valid these assumptions become. It explains why the analytic model 
works better at $R_{f}>20\,h^{-1}$Mpc. 
 
\subsection{Effect of the Cosmic Web}\label{sec:shape_web}

Now, we would like to investigate whether or not the strength of the alignments between the shapes of the 
galactic halos and the tidal eigenvectors depend on the types of the cosmic web. 
Following the conventional scheme \citep{hah-etal07a}, we classify the galactic halos of each subsample into the 
knot, filament, sheet and void halos according to the signs of the eigenvalues of ${\bf T}$ at the grids where the halos are located: 
%%%%%%%%%%%%%%%%%%%%%%%%%%%%%%%%%%%%%%%%%%%%%%%%%%%%%%%%%%%%%%%
\begin{eqnarray}
\label{eqn:knot_con}
\lambda_{3}>0 &\rightarrow& {\rm knot}\, ,\\
\label{eqn:fil_con}
\lambda_{2}>0\, , \lambda_{3}<0 &\rightarrow& {\rm filament}\, , \\
\label{eqn:sheet_con}
\lambda_{1}>0\, , \lambda_{2}<0 &\rightarrow& {\rm sheet}\, , \\
\label{eqn:void_con}
\lambda_{1}<0 &\rightarrow& {\rm void}\, .
\end{eqnarray}
%%%%%%%%%%%%%%%%%%%%%%%%%%%%%%%%%%%%%%%%%%%%%%%%%%%%%%%%%%%%%%%

Using only those galactic halos embedded in the same type of the cosmic web, we redo the whole analysis
described in Section \ref{sec:shape_test}. Figures \ref{fig:pro_axis_knot}-\ref{fig:pro_axis_void} show the same as 
Figures \ref{fig:pro_axis} but only with the knot, filament, sheet and void halos, respectively, showing how the shape 
alignment depends on the web environment. 
Figure \ref{fig:dt_web} plots $\langle d_{t}\rangle$ versus $M$ for the four different cases of the web type. 
As can be seen, the value of $\langle d_{t}\rangle$ increases more sharply with the increment of $M$ for the cases of the sheet and 
void halos than for the cases of the knot and filament counterparts, which indicates that the shapes of the galactic halos in the relatively 
low-density regions tend to be more strongly aligned with those in the relatively high-density regions. 
Given that the galactic halos located in the knot and filament regions are expected to have formed earlier and undergone more 
severe nonlinear evolutions than those in the sheet and void regions \citep{GW07}, the results shown in Figure \ref{fig:dt_m} imply 
that the nonlinear evolution in denser environments will play a decisive role in diminishing the strength of the tidally induced shape 
alignments of the galactic halos. 

It is interesting to note that the results shown in Figures \ref{fig:pro_axis_knot}-\ref{fig:dt_web} are in direct contradiction with that of \citet{xia-etal17} who found the strongest shape alignments of the halos 
in the knot environments. We think that this apparent inconsistency between our and their results may be related 
the difference in the web classification scheme. In their analysis, the types of the cosmic web are classified according to the 
signs of the eigenvalues of the Hessian matrix of the density field. Whereas in our analysis the eigenvalues of the Hessian 
matrix of the gravitational potential field (i.e, tidal field) are used for the web classification. 

%It is, however, not only the high density that drives the non-linear evolution. The central sections 
%of the voids, in spite of its very low density, have been known to be also deeply in the nonlinear regime \citep{SW04}.  
%In fact, among the four types of the cosmic web, it is only the sheets that are still in the quasi-linear regime \citep[e.g.,][]{bri-etal16}, 
%which explains why the sheet galactic halos show stronger shape alignments than the void counterparts in the low and medium mass 
%ranges. In the high-mass range, however, the void and sheet galactic halos exhibit similar strengths of the shape alignments. Our 
%explanation for this is as follows. The high-mass void galactic halos are usually located not in the central sections of the voids but in their 
%outskirts wrapped by the sheets and thus should be in close proximity to the neighbor sheet galactic halos that are in quasi-linear 
%regime.  

Figures \ref{fig:pro_axis_filter_knot}-\ref{fig:pro_axis_filter_void} show the same as Figures \ref{fig:pro_axis_filter} but with only 
those high-mass galactic halos located in the knot, filament, sheet and void environments, respectively. Figure \ref{fig:dt_webf} 
plots $\langle d_{t}\rangle$ versus $R_{f}$ for the four different cases of the web type. 
The decrement of the alignment strength with the increment of $R_{f}$ is found for all of the four types of the 
cosmic web. The void (knot) galactic halos show the most (least) rapid change of $\langle d_{t}\rangle$ with $R_{f}$. 
The web-dependence of the rate of the change of $\langle d_{t}\rangle$ with $R_{f}$ shown in Figure \ref{fig:dt_webf} implies that the 
strength of the tidally induced shape alignments of the galactic halos is determined not only by the difference between 
$R_{g}$ and $R_{f}$ but also by the strength of the cross correlations between the tidal fields smoothed on different scales. 
In the denser knot and filament environments, although the nonlinearity diminishes the strength of the initially induced shape alignments 
with the large-scale tidal fields, the stronger cross correlations between the tidal fields smoothed on different scales slow down the rate of 
the decrement of the strength of the shape alignments with the increment of $R_{f}$. Whereas, in the less dense sheet and void regions 
where the strongest signals of the shape alignments are found on the scale of $R_{f}=5\,h^{-1}$Mpc, the weaker cross-correlations 
between the tidal fields on different scales cause the strengths of the shape alignments to decrease quite rapidly as $R_{f}$ increases. 

Figures \ref{fig:pro_axis_knot}-\ref{fig:dt_webf} clearly demonstrate that our analytic model with the empirically determined parameter, 
Equations (\ref{eqn:hpro_axis_dt})-(\ref{eqn:dt_sol}), makes a quantitative success in 
describing simultaneously and consistently the amplitudes and behaviors of the three probability density functions, 
$\{p(\vert\hat{\bf u}_{i}\cdot\hat{\bf e}\vert)\}_{i=1}^{3}$, 
for all of the cases of the galactic mass ranges $M$, the smoothing scales $R_{f}$ and the types of the cosmic web.  
This notable success of our analytic model confirms the validity of the key assumption made for Equation (\ref{eqn:eiej}) 
that the covariances of the shapes of the galactic halos have a linear dependence on the large-scale tidal fields 
\citep{cat-etal01,LP08,HZ08}. 

\section{Tidally Induced Spin Alignments}\label{sec:spin}

\subsection{Analytic Models}\label{sec:spin_model}

Employing the analytic model based on the LTT theory developed by \citet{LP00,LP01}, \citet{lee04} derived the probability density 
functions of the coordinates of the unit spin vectors, $\hat{\bf s}$, given $\hat{\bf T}$ \citep[see also][]{lee-etal18}:
%%%%%%%%%%%%%%%%%%%%%%%%%%%%%%%%%%%%%%%%%%%%%%%%%%%%%%%%%%%%
\begin{equation}
\label{eqn:model1}
p(\vert\hat{\bf u}_{i}\cdot\hat{\bf s}\vert)= \frac{1}{2\pi}\int_{0}^{2\pi}\,
\left[\prod_{n=1}^{3}\left(1+c_{t}-3c_{t}\hat{\lambda}^{2}_{n}\right)\right]^{-\frac{1}{2}}
\left(\sum_{l=1}^{3}\frac{\vert\hat{\bf u}_{l}\cdot\hat{\bf s}\vert}{1+c_{t}-3c_{t}\hat{\lambda}^{2}_{l}} \right)^{-\frac{3}{2}}\,d\phi_{jk}\, , 
\end{equation}
%%%%%%%%%%%%%%%%%%%%%%%%%%%%%%%%%%%%%%%%%%%%%%%%%%%%%%%%%%%
where $c_{t}$ is the spin correlation parameter in the range of $[0,1]$ \citep{LP01}. The larger value of $c_{t}$ is translated 
into the stronger $\hat{\bf u}_{2}$-$\hat{\bf s}$ alignment. 
Although Equation (\ref{eqn:model1}) is quite similar to Equation (\ref{eqn:hpro_axis_dt}), there is an obvious difference: 
the former is expressed in terms of $\hat{\lambda}^{2}_{i}$, while the latter in terms of $\hat{\lambda}_{i}$. 
This difference originates from the fact that the covariances of the spin vectors of the galactic halos have a quadratic dependence on 
$\hat{\bf T}$ according to the LTT theory \citep{dor70,whi84}. 

The core assumption that underlies Equation (\ref{eqn:model1}) is that the rescaled covariance, 
$\langle{s}_{i}{s}_{j}\vert\hat{\bf T}\rangle$, can be written as \citep{LP00}
%%%%%%%%%%%%%%%%%%%%%%%%%%%%%%%%%%%%%%%%%%%%%%%%%%%%%%%%%%%%
\begin{equation}
\label{eqn:ct}
\langle{s}_{i}{s}_{j}\vert\hat{\bf T}\rangle = \frac{1+c_{t}}{3}\delta_{ij} - c_{t}\sum_{k=1}^{3}\hat{T}_{ik}\hat{T}_{kj}\, . 
\end{equation}
%%%%%%%%%%%%%%%%%%%%%%%%%%%%%%%%%%%%%%%%%%%%%%%%%%%%%%%%%%%%
Solving Equation (\ref{eqn:ct}) for $c_{t}$ in the principal frame of $\hat{\bf T}$ gives
\footnote{In \citet{lee-etal18}, there was a typo in the formula. It is corrected here.} \citep{LP01}
%%%%%%%%%%%%%%%%%%%%%%%%%%%%%%%%%%%%%%%%%%%%%%%%%%%%%%%%%%%%
\begin{equation}
\label{eqn:ct_sol}
c_{t} = \frac{10}{3} - 10\sum_{i=1}^{3}\hat{\lambda}_{i}^{2}\langle\hat{s}^{2}_{i}\vert\hat{\bf T}\rangle\, . 
\end{equation}
%%%%%%%%%%%%%%%%%%%%%%%%%%%%%%%%%%%%%%%%%%%%%%%%%%%%%%%%%%%%
Equation (\ref{eqn:ct_sol}) enables us to evaluate the value of $c_{t}$ directly from the values of $\hat{\bf s}$, 
$\{\hat{\lambda}_{i}\}_{i=1}^{3}$ and $\{\hat{\bf u}_{i}\}_{i=1}^{3}$.  Several observational and numerical studies showed that 
this analytic model, Equations (\ref{eqn:model1})-(\ref{eqn:ct_sol}), was indeed useful and adequate in describing the tidally 
induced spin alignments especially in the sheet environments  \citep[e.g.,][]{nav-etal04,tru-etal06,LE07,lee-etal18}. 
As mentioned in Section \ref{sec:intro}, however, the LTT theory breaks down on the mass scale below 
$M_{\rm t}\sim 10^{12}\,h^{-1}\,M_{\odot}$. The numerical analyses based on recent large high-resolution $N$-body simulations 
found that the spin flip, a transition of the tendency from the $\hat{\bf u}_{2}$-$\hat{\bf s}$ ($M> M_{\rm t}$) alignments to the 
$\hat{\bf u}_{3}$-$\hat{\bf s}$ alignments ($M\le M_{\rm t}$) occurs \citep{ara-etal07,cod-etal12,vee-etal18} and that the value of the 
transition mass scale, $M_{\rm t}$, depends on the type of the cosmic web \citep{lib-etal13}.

Now, we would like to construct a new model that might describe quantitatively the transition of the spin alignment tendency at 
$M_{\rm t}$ and its dependence on the type of the cosmic web. In the light of the previous studies which claimed that the 
nset of the non-Gaussianity of the tidal fields even on large scales would cause the covariance, 
$\langle{s}_{i}{s}_{j}\vert\hat{\bf T}\rangle$, to scale linearly with $\hat{\bf T}$ \citep{HZ08,LP08}, we first modify 
Equation (\ref{eqn:ct}) into 
%%%%%%%%%%%%%%%%%%%%%%%%%%%%%%%%%%%%%%%%%%%%%%%%%%%%%%%%%%%%
\begin{equation}
\label{eqn:cdt}
\langle{s}_{i}{s}_{j}\vert\hat{\bf T}\rangle =  \frac{(1+c_{t}+d_{t})}{3}\delta_{ij} - 
c_{t}\sum_{k=1}^{3}\hat{T}_{ik}\hat{T}_{kj} - d_{t}\hat{T}_{ij}\,  , 
\end{equation}
%%%%%%%%%%%%%%%%%%%%%%%%%%%%%%%%%%%%%%%%%%%%%%%%%%%%%%%%%%%%
where two spin correlation parameters, $c_{t}$ and $d_{t}$, both lying in the range of $[0,1]$, are introduced to 
correlate $\hat{\bf s}$ to $\hat{\bf u}_{2}$ and to $\hat{\bf u}_{3}$, respectively.  
If the first spin correlation parameter, $c_{t}$, is close to zero and the second spin correlation parameter, $d_{t}$, is close to unity, 
then the spin vectors $\hat{\bf s}$ will show strong alignments with $\hat{\bf u}_{3}$ just like the shape vectors, $\hat{\bf e}$. 
If $c_{t} $ is close to unity and $d_{t}$ is close to zero, then it will be reduced to the original model, Equation (\ref{eqn:model1}), 
which describes the $\hat{\bf u}_{2}$-$\hat{\bf s}$ alignments. If both of the parameters are close to zero, then the spin vectors of the 
galactic halos will be random having no correlations with the large-scale tidal fields. 

Replacing Equation (\ref{eqn:ct}) by Equations (\ref{eqn:cdt}) in the original derivation of Equation (\ref{eqn:model1}), it is 
straightforward to show that the probability density functions, $p(\vert\hat{\bf u}_{i}\cdot\hat{\bf s}\vert)$, can be expressed as 
%%%%%%%%%%%%%%%%%%%%%%%%%%%%%%%%%%%%%%%%%%%%%%%%%%%%%%%%%%%%
\begin{eqnarray}
\label{eqn:model3}
p(\vert\hat{\bf u}_{i}\cdot\hat{\bf s}\vert)&=& \frac{1}{2\pi}\int_{0}^{2\pi}\,
\left[\prod_{n=1}^{3}\left(1+c_{t}-3c_{t}\hat{\lambda}^{2}_{n}+d_{t}-3d_{t}\hat{\lambda}_{n}\right)\right]^{-\frac{1}{2}}\times \, 
\nonumber \\
&&\left[\sum_{l=1}^{3}\left(\frac{\vert\hat{\bf u}_{l}\cdot{\bf s}\vert}
{1+c_{t}-3c_{t}\hat{\lambda}^{2}_{l}+d_{t}-3d_{t}\hat{\lambda}_{l}}\right)\right]^{-\frac{3}{2}}\,d\phi_{jk}\, .
\end{eqnarray}
%%%%%%%%%%%%%%%%%%%%%%%%%%%%%%%%%%%%%%%%%%%%%%%%%%%%%%%%%%%
Equation (\ref{eqn:ct_sol}), which was originally derived in the LTT theory, holds true even when the covariance, 
$\langle\hat{s}_{i}\hat{s}_{j}\vert\hat{\bf T}\rangle$, has an additional term, 
since the second and third terms in Equation (\ref{eqn:cdt}) are uncorrelated due to 
$\langle\hat{T}_{ik}\hat{T}_{kl}\hat{T}_{lj}\rangle=0$ \citep[see Appdendix E in][]{LP01}.
Thus, the same formula as Equation (\ref{eqn:ct_sol}) can be used to obtain the value of $c_{t}$ for this new model. 
Likewise, the same formula as Equation (\ref{eqn:dt_sol}) but with $\hat{\bf e}$ replaced by $\hat{\bf s}$ 
can be used to obtain the value $d_{t}$ as 
%%%%%%%%%%%%%%%%%%%%%%%%%%%%%%%%%%%%%%%%%%%%%%%%%%%%%%%%%%%%%
\begin{equation}
\label{eqn:dt_spin}
d_{t} = -\frac{5}{3}\sum_{i=1}^{3}\hat{\lambda}_{i}\langle\hat{s}^{2}_{i}\vert\hat{\bf T}\rangle\, .
\end{equation}
%%%%%%%%%%%%%%%%%%%%%%%%%%%%%%%%%%%%%%%%%%%%%%%%%%%%%%%%%%%%%%%

In Section \ref{sec:spin_test}, we will numerically test three models for the galaxy spin alignments, 
\texttt{model I}, \texttt{model II} and \texttt{model III}. The \texttt{model III} is Equation (\ref{eqn:model3}) with two non-zero 
parameters, $c_{t}$ and $d_{t}$. The \texttt{model I} is Equation (\ref{eqn:model3}) with $d_{t}=0$. It is identical to the original 
model based on the LTT theory, Equation (\ref{eqn:model1}). The \texttt{model II} is Equation (\ref{eqn:model3}) with $c_{t}=0$. 
It has the same functional form as Equation (\ref{eqn:hpro_axis_dt}) for the tidally induced shape alignments. 
 
\subsection{Numerical Tests}\label{sec:spin_test}

To numerically obtain three probability density functions, $\{p(\vert\hat{\bf u}_{i}\cdot\hat{\bf s}\vert)\}_{i=1}^{3}$, 
we perform the exactly same calculations as presented in Section \ref{sec:shape_test}, but with $\hat{\bf e}$ replaced by 
$\hat{\bf s}$. For the evaluation of the three analytic models, we first determine the ensemble values of $\langle c_{t}\rangle$ 
and $\langle d_{t}\rangle$ for each subsample as, 
%%%%%%%%%%%%%%%%%%%%%%%%%%%%%%%%%%%%%%%%%%%%%%%%%%%%%%%%%%%%%%%
\begin{eqnarray}
\label{eqn:cdt_sol}
\langle c_{t}\rangle &=&  \frac{1}{N_{t}}\sum_{\alpha=1}^{N_{t}}
\left[\frac{10}{3}-10\sum_{i=1}^{3}\hat{\lambda}^{2}_{\alpha,i}\vert\hat{\bf u}_{\alpha,i}\cdot\hat{\bf s}_{\alpha}\vert^{2}\right]\, ,\\
\label{eqn:mdt_spin}
\langle d_{t}\rangle &=& \frac{1}{N_{t}}\sum_{\alpha=1}^{N_{t}}\left[-\frac{5}{3}
\sum_{i=1}^{3}\hat{\lambda}_{\alpha,i}\vert\hat{\bf u}_{\alpha,i}\cdot\hat{\bf s}_{\alpha}\vert^{2}\right]\,  ,
\end{eqnarray}
%%%%%%%%%%%%%%%%%%%%%%%%%%%%%%%%%%%%%%%%%%%%%%%%%%%%%%%%%%%%%%%
and put these ensemble average values into Equation (\ref{eqn:model3}) to evaluate the \texttt{model III}. Putting 
$\langle c_{t}\rangle$ ($\langle d_{t}\rangle$) into Equation (\ref{eqn:model3}) and setting 
$\langle d_{t}\rangle$ ($\langle c_{t}\rangle$) at zero, we evaluate the \texttt{mode I} (\texttt{model II}).

Figure \ref{fig:pro_spin} plots the numerically obtained probability density functions, $\{p(\vert\hat{\bf u}_{i}\cdot\hat{\bf s}\vert)\}_{i=1}^{3}$ 
(filled dots), and compares them with the \texttt{model I} (blue lines), \texttt{model II} (green lines) and \texttt{model III} (red lines). 
As can be seen, the three functions, $\{p(\vert\hat{\bf u}_{i}\cdot\hat{\bf s}\vert)\}_{i=1}^{3}$, have much lower amplitudes than 
$\{p(\vert\hat{\bf u}_{i}\cdot\hat{\bf e}\vert)\}_{i=1}^{3}$ displayed in Figure \ref{fig:pro_axis}. It indicates that the spin vectors of the 
galactic halos are much less strongly aligned with the large-scale tidal fields than the shape vectors, which is 
consistent with the results of the previous numerical and observational studies \citep[e.g.,][]{hah-etal07b,for-etal14,zha-etal15}. 

The occurrence of the spin-flip phenomenon is indeed witnessed: For the case of the lower mass galactic halos 
with $M< 10^{12}\,h^{-1}M_{\odot}$, the unit spin vectors, $\hat{\bf s}$, tend to be aligned not with the intermediate 
eigenvectors, $\hat{\bf u}_{2}$, but with the minor eigenvectors, $\hat{\bf u}_{3}$, while the high-mass galactic halos 
with $M\ge 10^{12}\,h^{-1}M_{\odot}$, exhibit the stronger alignments of $\hat{\bf s}$ with $\hat{\bf u}_{2}$ rather than with 
$\hat{\bf u}_{3}$, which is quite consistent with the previous numerical results 
\citep[e.g.,][]{ara-etal07,hah-etal07b,cod-etal12,lib-etal13,dub-etal14,che-etal16,vee-etal18}. 

The strength of the $\hat{\bf u}_{3}$-$\hat{\bf s}$ alignments tends to decrease with $M$, while the strengths of the 
$\hat{\bf u}_{2}$-$\hat{\bf s}$ alignments increases with $M$. These opposite trends can be quantitatively described by the 
variation of the first and second spin correlation parameters with $M$ as shown in the top and bottom panels of 
Figure \ref{fig:cdt_spin_m}, respectively.  As can be seen, $\langle d_{t}\rangle$ is larger than $\langle c_{t}\rangle$ in the lower mass 
range of $M<10^{12}\,h^{-1}M_{\odot}$ but drops below $\langle c_{t}\rangle$ in the higher mass range of 
$M\ge 10^{12}\,h^{-1}M_{\odot}$. The transition mass scale of the spin-flip corresponds to the moment when 
$\langle d_{t}\rangle$ becomes lower than $\langle c_{t}\rangle$. 

For the case of the lowest and low-mass galactic halos with $M<5\times 10^{11}\,h^{-1}M_{\odot}$, both of the models \texttt{II} and 
\texttt{III} succeed in matching simultaneously the amplitudes and behaviors of the three numerically obtained probability density functions. 
The \texttt{model II} is almost identical to the \texttt{model III} in these low-mass ranges, since the values of $\langle c_{t}\rangle$ 
obtained via Equation (\ref{eqn:cdt_sol}) are  low for these cases. 
It is also worth noting that the signal of the strong $\hat{\bf u}_{1}$-$\hat{\bf s}$ anti-alignments is found to increase with $M$ 
whose behavior is well described by both of the model \texttt{II} and \texttt{III}. 
The success of the \texttt{model II} and \texttt{model III} and the failure of the \texttt{model I} in describing the amplitudes and 
behaviors of $p(\vert\hat{\bf u}_{3}\cdot\hat{\bf s}\vert)$ and $p(\vert\hat{\bf u}_{1}\cdot\hat{\bf s}\vert)$ are also found for the 
case of the medium-mass halos (second from the bottom panels in Figure \ref{fig:pro_spin}). 

It is, however, interesting to note that in this medium-mass range the spin vectors, $\hat{\bf s}$, exhibit a weak but non-negligible 
alignment with the intermediate eigenvectors, $\hat{\bf u}_{2}$, which tendency is  properly described by both of \texttt{model I} and 
\texttt{model III} but not by the \texttt{model II}. 
For the case of the high-mass galactic halos, the unit spin vectors, $\hat{\bf s}$, turn out to be more strongly aligned with 
$\hat{\bf u}_{2}$ than with $\hat{\bf u}_{3}$, which cannot be described by the \texttt{model II}. But, the alignments of $\hat{\bf s}$ 
with $\hat{\bf u}_{3}$ and its anti-alignments with $\hat{\bf u}_{1}$ are still well described by the \texttt{model II} and \texttt{III} but 
not by the \texttt{model I}. Thus, it is only the \texttt{model III} that agrees concurrently and consistently with the numerically obtained 
three probability density functions,  $p(\vert\hat{\bf u}_{3}\cdot\hat{\bf s}\vert)$, $p(\vert\hat{\bf u}_{2}\cdot\hat{\bf s}\vert)$, and 
$p(\vert\hat{\bf u}_{1}\cdot\hat{\bf s}\vert)$, in all of the four mass ranges. 

Figure \ref{fig:pro_spin_filter} shows the same as the bottom panels of Figure \ref{fig:pro_spin} but for the cases that the 
tidal fields are smoothed on three larger scales of $R_{f}=10,\ 20$ and $30\,h^{-1}$Mpc, in the top, middle and bottom panels, 
respectively. As can be seen, the increment of $R_{f}$ decreases the alignment strengths even more 
rapidly than for the case of the shape alignments.  Note that it is only the \texttt{model III} that succeeds in making good 
simultaneous descriptions of the amplitudes and behaviors of  $\{p(\vert\hat{\bf u}_{i}\cdot\hat{\bf s}\vert)\}_{i=1}^{3}$ 
for all of the three cases of $R_{f}$. 

Although the \texttt{model III} achieves overall good agreements with the numerical results, some discrepancies between its description  
and the numerical results are found. As can be seen in Figures \ref{fig:pro_spin}-\ref{fig:pro_spin_filter}, the numerically obtained three 
probability functions display substantially fluctuating behaviors especially for the case of the high-mass galactic halos. However, the 
increment of $R_{f}$ reduces these discrepancies as shown in Figure \ref{fig:pro_spin_filter}, which implies that the inaccuracies 
associated with the approximations of ${\bf T}$ as a Gaussian random field and $p({\bf s}\vert{\bf T})$ as a multivariate Gaussian 
distribution in the derivation of Equation (\ref{eqn:model3}) should be largely responsible for these discrepancies. 

The uncertainties involved in the measurements of the spin vectors of the galactic halos may be another source of the discrepancies. 
Since the spin direction of a galactic halo is dominantly determined by the positions and velocities of the outmost DM particles from 
the halo center, its measurement would depend sensitively on the dynamical state of the galactic halo, halo-finding algorithm and 
definition of the virial radius. If a high-mass galactic halo has yet to be fully relaxed and/or in the middle of merging, containing 
multiple substructures, the measurement of its spin direction is likely to suffer from substantial uncertainties, which in turn 
would cause mismatches between the analytical and the numerical results on the spin alignments with the large-scale tidal field.     

Figure \ref{fig:cdt_spin_f} shows how the first and second spin parameters vary with $R_{f}$ for the high-mass galactic halos in the 
top and bottom panels, respectively. As can be seen,  both of the parameters decrease with the increment of $R_{f}$. 
The two parameters, however, show different variations with $M$. The first spin parameter, $\langle c_{t}\rangle$, decreases more rapidly 
with the increment of $R_{f}$ than the second spin parameter, $\langle d_{t}\rangle$. 
It is found that $\langle c_{t}\rangle>\langle d_{t}\rangle$ at $R_{f}\le 20\,h^{-1}$Mpc, while $c_{t}<d_{t}$ at $R_{f}=30\,h^{-1}$Mpc.  
This result implies that the occurrence of the spin flip phenomenon is contingent on the sizes of 
the large-scale structures.  Suppose that the galaxies with masses in the range of $0.5\le M/(10^{11}\,h^{-1}\,M_{\odot})\le 50$ embedded 
in a coherent large-scale structure like a filament with size $R_{f}\ge 30\,h^{-1}$Mpc. According to our results, the spin vectors of those 
galaxies would not flip, with their spins always aligned with the elongated axes of the host filament since 
$\langle d_{t}\rangle$ is always higher than $\langle c_{t}\rangle$ in the given mass range (see Section \ref{sec:spin_web}). 

\subsection{Effect of the Cosmic Web}\label{sec:spin_web}

Following the same procedure as presented in Section \ref{sec:shape_web}, we investigate how the probability density functions, 
$\{p(\vert\hat{\bf u}_{i}\cdot\hat{\bf s}\vert)\}_{i=1}^{3}$, depend on the type of the cosmic web. 
Figures \ref{fig:pro_spin_knot}-\ref{fig:pro_spin_void} show the same as Figure \ref{fig:pro_spin} but only with the galactic 
halos located in the knot, filament, sheet and void environments, respectively. 
In the knot environments (Figure \ref{fig:pro_spin_knot}), the unit spin vectors, $\hat{\bf s}$, of the galactic halos are found strongly 
aligned with the minor eigenvector $\hat{\bf u}_{3}$ in all of the four mass ranges (i.e., no spin-flip). For the cases of the lowest-mass, 
low-mass and medium-mass knot galactic halos, we find $\hat{\bf s}$ to be slightly anti-aligned rather than aligned with $\hat{\bf u}_{2}$, 
while the high-mass knot galactic halos show weak $\hat{\bf u}_{2}$-$\hat{\bf s}$ alignments. Both of the \texttt{model II} and 
\texttt{model III} describe well the $\hat{\bf u}_{3}$-$\hat{\bf s}$ alignment and the $\hat{\bf u}_{1}$-$\hat{\bf s}$ anti-alignment. 
However, \texttt{model II} cannot describe the observed  tendency of the $\hat{\bf u}_{2}$-$\hat{\bf s}$ anti-alignment in the mass 
scale of $5\le M/(10^{11}\,h^{-1}\,M_{\odot})< 10$ while the \texttt{model III} can. 
It is interesting to see that the \texttt{model I} describes better the observed $\hat{\bf u}_{2}$-$\hat{\bf s}$ anti-alignments in 
the medium-mass range better than the \texttt{model II} although it still notoriously fails in describing the 
observed strong $\hat{\bf u}_{3}$-$\hat{\bf s}$ alignments and $\hat{\bf u}_{1}$-$\hat{\bf s}$ anti-alignments. 

The filament galactic halos yield much stronger $\hat{\bf u}_{3}$-$\hat{\bf s}$ alignment and $\hat{\bf u}_{1}$-$\hat{\bf s}$ anti-alignment 
in all of the four mass ranges than the knot counterpart, although the behaviors of $\{p(\vert\hat{\bf u}_{i}\cdot\hat{\bf s}\vert)\}_{i=1}^{3}$ 
between the two cases are quite similar to each other (Figure \ref{fig:pro_spin_fil}).  The high-mass filament galactic halos 
show a substantial $\hat{\bf u}_{2}$-$\hat{\bf s}$ alignment whose strength is comparable to that of the $\hat{\bf u}_{3}$-$\hat{\bf s}$
(i.e., the occurrence of the spin flip).  Although the \texttt{model III} works quite well in matching the numerically obtained probability 
density functions, it is interesting to note that the \texttt{model II} gives a better description of $p(\vert\hat{\bf u}_{2}\cdot\hat{\bf s}\vert)$ 
than the \texttt{model III} in the mass range of $M< 10^{12}\,h^{-1}M_{\odot}$. 

The sheet galactic halos exhibit a different trend (Figure \ref{fig:pro_spin_sheet}). Their spin vectors tend to  
lie in the plane spanned by $\hat{\bf u}_{2}$ and $\hat{\bf u}_{3}$, being orthogonal to $\hat{\bf u}_{1}$. The increment of $M$ 
leads to the stronger $\hat{\bf u}_{2}$-$\hat{\bf s}$ alignment and $\hat{\bf u}_{1}$-$\hat{\bf s}$ anti-alignment but weaker 
$\hat{\bf u}_{3}$-$\hat{\bf s}$ alignment. For the case of the lowest-mass and low-mass sheet galactic halos, 
the $\hat{\bf u}_{2}$-$\hat{\bf s}$ alignment tendency is weaker than the $\hat{\bf u}_{3}$-$\hat{\bf s}$ alignment.  
For the case of the medium-mass sheet galactic halos, the $\hat{\bf u}_{2}$-$\hat{\bf s}$ alignment begins to exceed in strength the 
$\hat{\bf u}_{3}$-$\hat{\bf s}$ alignment (i.e., occurrence of the spin flip). The strongest signal of the  $\hat{\bf u}_{2}$-$\hat{\bf s}$ 
alignments is found from the high-mass sheet galactic halos, which result is consistent with the previous numerical finding of 
\citet{hah-etal07a}. 
As can be seen, only the \texttt{Model III} succeeds in describing simultaneously and consistently the behaviors of 
$\{p(\vert\hat{\bf u}_{i}\cdot\hat{\bf s}\vert)\}_{i=1}^{3}$, fairly well for the case of the sheet galactic halos in all of the 
four mass ranges. 

This result is inconsistent with the observational finding of \citet{zha-etal15} that the galaxies in the knot environments exhibited 
the strongest spin alignments with the tidal fields.  We suspect  that two factors may have caused this inconsistency between 
the numerical and observational results on the web dependence of the spin alignments. First, the difference in the way 
in which the tidal fields were constructed. In the work of \citet{zha-etal15}, the tidal fields, $\hat{\bf T}$, were constructed from the spatial 
distributions of the galaxy groups, while in the SMDPL the spatial distribution of the DM particles were used. 
Second, the difference in the measurements of $\hat{\bf s}$: In the observational analysis of \citet{zha-etal15}, the unit 
spin vectors $\hat{\bf s}$ were determined from the luminous parts of the galaxies while in the current numerical analyses, all of the 
constituent DM particles determine $\hat{\bf s}$. 

The weakest spin alignments with the large-scale tidal fields are found in the void environments (Figure \ref{fig:pro_spin_void}). 
Although the signals are quite lower than those yielded by the sheet galactic halos, 
the behaviors of $\{p(\vert\hat{\bf u}_{i}\cdot\hat{\bf s}\vert)\}_{i=1}^{3}$ obtained from the void galactic halos are quite similar to 
those from the sheet galactic halos:  the alignments of $\hat{\bf s}$ with $\hat{\bf u}_{2}$ and $\hat{\bf u}_{3}$. 
The former (the latter) alignment become stronger (weaker) with the increment of $M$. 
For the lowest-mass and low-mass void galactic halos (top two panels), the $\hat{\bf u}_{3}$-$\hat{\bf s}$ alignment is slightly stronger 
than the $\hat{\bf u}_{2}$-$\hat{\bf s}$ alignment. Only the \texttt{Model III} pulls it off to describe simultaneously the behaviors of 
$\{p(\vert\hat{\bf u}_{i}\cdot\hat{\bf s}\vert)\}_{i=1}^{3}$.  For the case of the medium-mass and high-mass void galactic halos, however, 
the large errors make it difficult to interpret the numerical results and to make a fair comparison of them with the three models. 

Figures \ref{fig:ct_spin_web} and \ref{fig:dt_spin_web} plot $\langle c_{t}\rangle$ and $\langle d_{t}\rangle$ 
versus $M$ for the four different web types, respectively.  Although the increment of the first spin correlation parameter, 
$\langle c_{t}\rangle$, with $M$ is universally shown, the increment rate sensitively depends on the web type. The most (least) rapid 
change of $\langle c_{t}\rangle$ with $M$ is found from the sheet (knot) galactic halos. Meanwhile, the second spin correlation parameter, 
$\langle d_{t}\rangle$, does not show strong variations with $M$. For the case of the high-mass filament and  void galactic halos, however, 
it shows an abrupt decrement with $M$.  

Defining the transition mass, $M_{t}$, as the one beyond which $\langle c_{t}\rangle$ exceeds $\langle d_{t}\rangle$, we expect the 
galactic halos with $M>M_{t}$ ($M\le M_{t}$) to exhibit the preferential $\hat{\bf u}_{2}$-$\hat{\bf s}$ ($\hat{\bf u}_{3}$-$\hat{\bf s}$) 
alignment. 
The results shown in Figures \ref{fig:ct_spin_web} and \ref{fig:dt_spin_web} imply that the value of $M_{t}$ depends on the web type. 
as shown in \citet{lib-etal13}. 
For the case of the knot galactic halos, no spin flip occurs in the given whole mass range since 
$\langle d_{t}\rangle$ is always larger than $\langle c_{t}\rangle$. The spin flip of the filament (sheet) galactic halos is expected 
to occur around $M_{t}\sim 5\times10^{12}\,h^{-1}M_{\odot}$ ($M_{t}\sim 10^{12}\,h^{-1}M_{\odot}$), while 
the void galactic halos show the lowest transition mass scale, $M_{t}\sim 5\times10^{11}\,h^{-1}M_{\odot})$. 

As done in Section \ref{sec:shape_web}, smoothing the tidal fields on three larger scales $R_{f}$ and repeating the whole calculation 
for each case of $R_{f}$, we investigate the dependence of the tendency and strength of the spin alignments on $R_{f}$ 
for the case of the high-mass galactic halos. 
Figures \ref{fig:pro_spin_filter_knot}-\ref{fig:pro_spin_filter_void} plot the same as the bottom panels of 
Figures \ref{fig:pro_spin_knot}-\ref{fig:pro_spin_void}, respectively,  but for the cases of $R_{f}=10,\ 20,\ 30\,h^{-1}$Mpc. 
As can be seen, whatever type of the cosmic web the galactic halos are embedded in, the increment of $R_{f}$ 
always decreases the alignment strength, which is well described by the \texttt{model III}.

For the cases of the high-mass galactic halos in the knot and filament regions, the increment of $R_{f}$ just decreases the strength of the 
spin alignments but does not change its tendency (Figures \ref{fig:pro_spin_filter_knot}-\ref{fig:pro_spin_filter_fil}). 
However, for the case of the high-mass sheet galactic halos (Figure \ref{fig:pro_spin_filter_sheet}), it changes both of the strength and the 
tendency of the spin alignments. On the scales of $R_{f}=10$ and $20\,h^{-1}$Mpc,  the high-mass sheet galactic halos show the stronger 
$\hat{\bf u}_{2}$-$\hat{\bf s}$ alignments than the $\hat{\bf u}_{3}$-$\hat{\bf s}$ alignments.  But, on the larger scale 
of $R_{f}=30\,h^{-1}$Mpc,  we witness a different tendency, the $\hat{\bf u}_{3}$-$\hat{\bf s}$ alignments seem slightly stronger 
than the $\hat{\bf u}_{2}$-$\hat{\bf s}$ alignments. In other words, if the sheet environment is defined on the scale equal to or larger than 
$30\,h^{-1}$Mpc, no spin-flip will occur in the given mass range.   Since both of $\hat{\bf u}_{2}$ and $\hat{\bf u}_{3}$ span the plane of a 
sheet \citep{zel70}, our result shown in Figure \ref{fig:pro_spin_filter_sheet} supports the claim of \citet{hah-etal07b} that the spin vectors 
of the DM halos have a universal tendency of lying in the plane of the sheet, regardless of the halo mass. 

\section{Summary and Discussion}\label{sec:con}

To study the large-scale tidal effect on the spin and shape orientations of the galaxies and the spin-flip phenomenon, 
we have considered three different analytic models, the \texttt{model I}, \texttt{model II} and \texttt{model III}.  The \texttt{model I}, 
Equation (\ref{eqn:model1}), which was originally developed by \citet{LP00} based on the LTT theory, describes the alignment 
tendency between the galaxy spin vectors, $\hat{\bf s}$, and the intermediate eigenvectors, $\hat{\bf u}_{2}$, of the large-scale tidal 
field, ${\bf T}$. 
The \texttt{model II}, Equation (\ref{eqn:hpro_axis_dt}), has been constructed here to describe the alignments 
(anti-alignments) of the galaxy shapes, $\hat{\bf e}$, with the minor (major) eigenvectors, $\hat{\bf u}_{3}$ ($\hat{\bf u}_{1}$) of ${\bf T}$. 
This model is based on the first order Lagrangian perturbation theory according to which the major principal axes of the inertia momentum 
tensors of the galactic halos are perfectly aligned with the minor principal axes of the local tidal tensors in the Lagrangian regime. 
The \texttt{model III}, Equation (\ref{eqn:model3}), is a practical formula constructed by combining the \texttt{model I} and 
\texttt{model II} to describe simultaneously the tidally induced shape and spin alignments. 

The \texttt{model I} (\texttt{model II}) carries a single parameter, $c_{t}$ ($d_{t}$),  which measures the 
strength of the alignment with $\hat{\bf u}_{2}$ ($\hat{\bf u}_{3}$). 
The \texttt{ model III} carries two parameters, $c_{t}$ and $d_{t}$, whose relative ratio determines the transition mass scale 
for the occurrence of the spin-flip. The first parameter, $c_{t}$, would reach the maximum value of unity, if the inertia momentum tensors 
of the galaxies are uncorrelated with the surrounding tidal tensors, while the second parameter, $d_{t}$, will attain the value of unity  if the 
two tensors are perfectly correlated.  These parameters can be empirically determined by Equation (\ref{eqn:cdt_sol}) directly from the 
measured values of $\hat{\bf e}$ and $\hat{\bf s}$ in the principal frame of $\hat{\bf T}$ without resorting to any fitting procedure.   

\begin{figure}
\begin{center}
\includegraphics[scale=1.0]{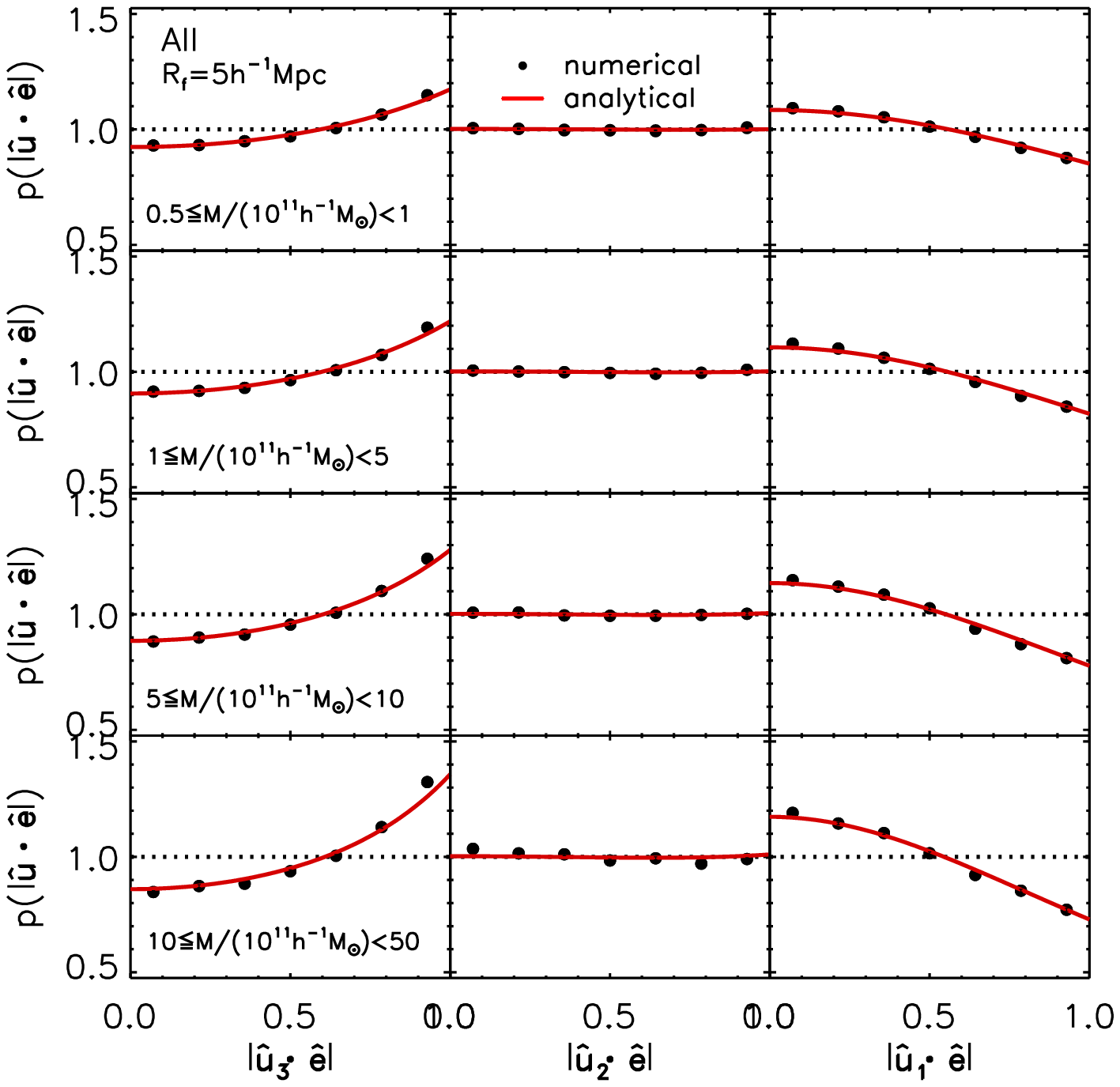}
\caption{Probability density distributions of three coordinates of the unit shape vectors, $\hat{\bf e}$, of 
the lowest-mass (top panel), low-mass (first middle panel), medium-mass (second middle panel) and 
high-mass (bottom panel) galactic halos  
in the principal frame spanned by the minor ($\hat{\bf u}_{3}$), intermediate ($\hat{\bf u}_{2}$), and 
major ($\hat{\bf u}_{1}$) eigenvectors of the tidal fields smoothed on the scale of $R_{f}=5\,h^{-1}$Mpc.  
In each panel, the numerical results are plotted as black filled circular dots with Poisson errors, while the analytic 
model, Equations (\ref{eqn:jpro_axis})-(\ref{eqn:hjpro_axis_dt}), with the empirically determined value of $d_{t}$ 
is shown as red solid line. The uniform constant probability density is depicted as black dotted line.}
\label{fig:pro_axis}
\end{center}
\end{figure}

\begin{figure}
\begin{center}
\includegraphics[scale=1.0]{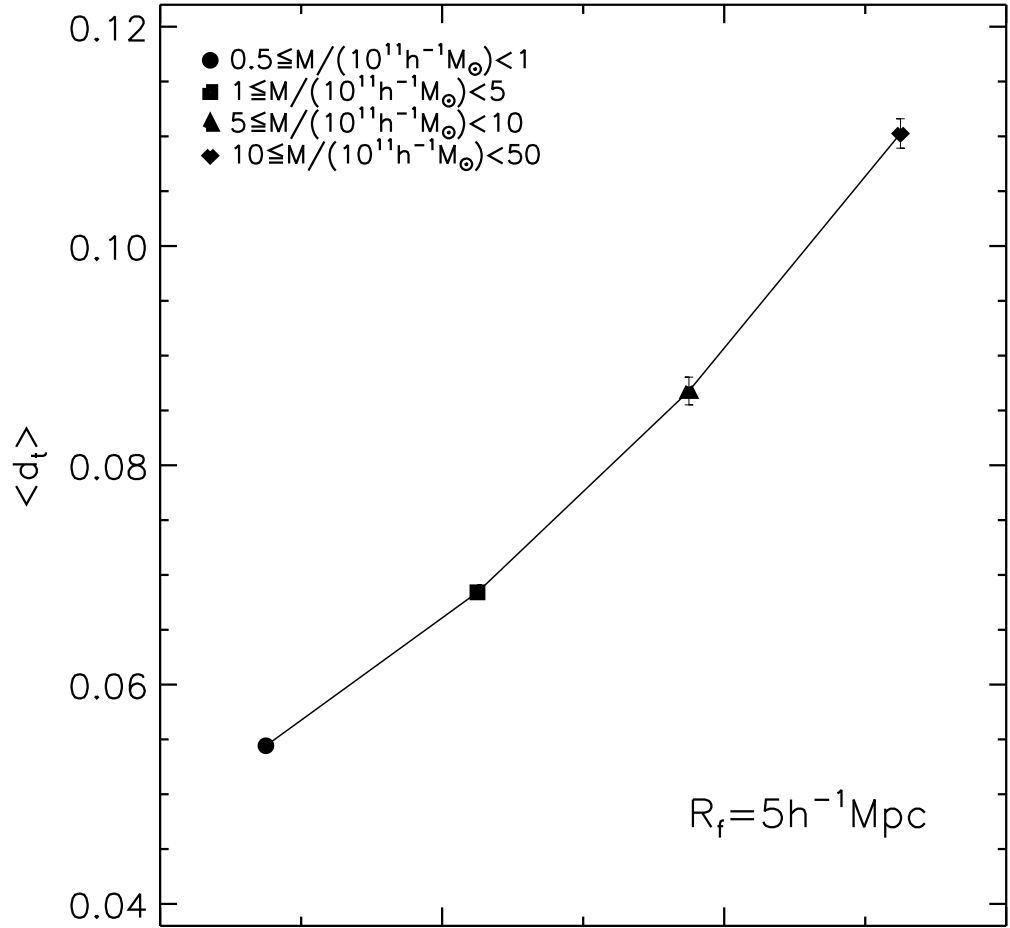}
\caption{Mean values of the shape correlation parameter $d_{t}$ averaged over the galactic halos 
belonging to four different mass ranges when the local tidal fields are smoothed on the scale of 
$R_{f}=5\,h^{-1}$Mpc.}
\label{fig:dt_m}
\end{center}
\end{figure}

\begin{figure}
\begin{center}
\includegraphics[scale=1.0]{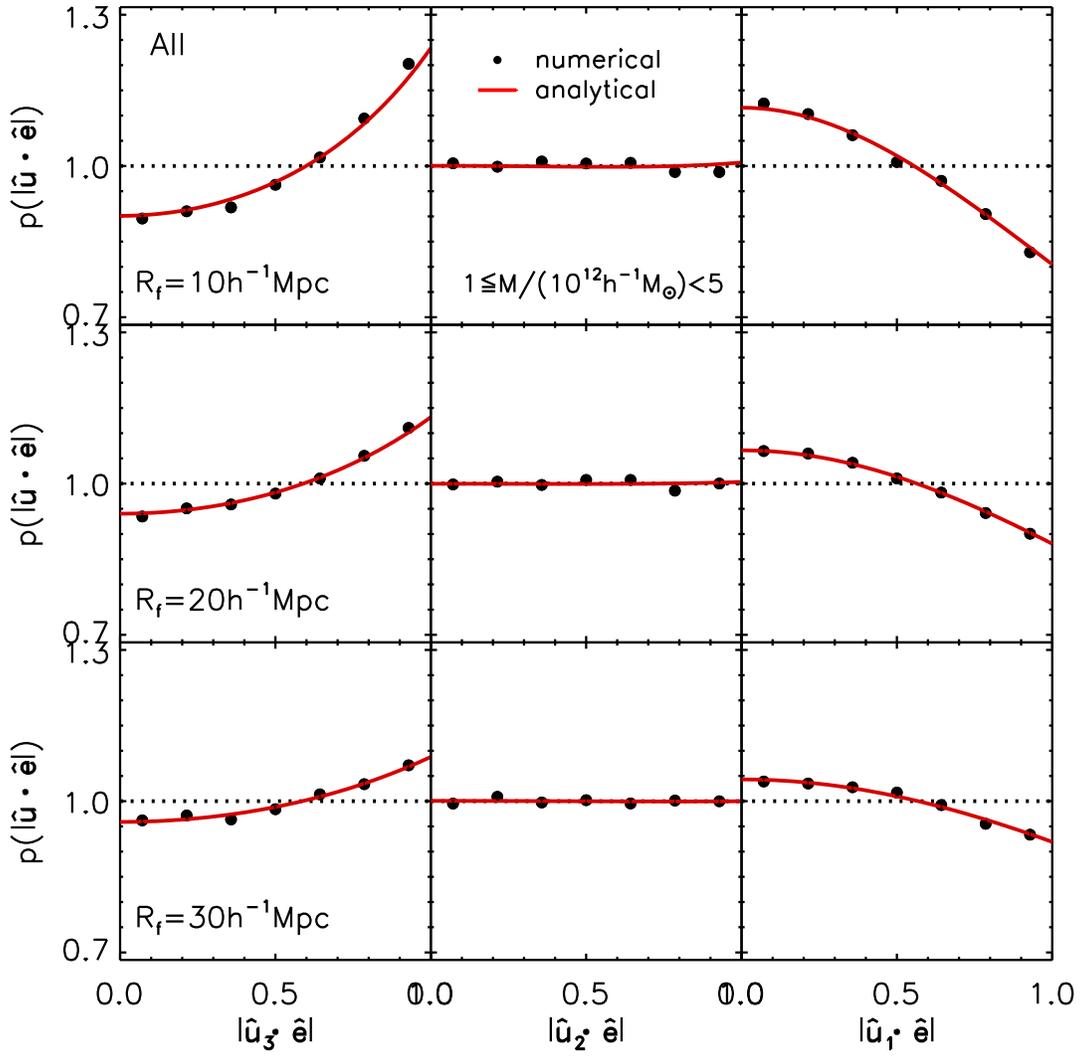}
\caption{Same as the bottom panels of Figure \ref{fig:pro_axis} but for the cases of three larger smoothing scales  
of $R_{f}=10,\ 20,\ 30\, h^{-1}$Mpc.}
\label{fig:pro_axis_filter}
\end{center}
\end{figure}

\begin{figure}
\begin{center}
\includegraphics[scale=1.0]{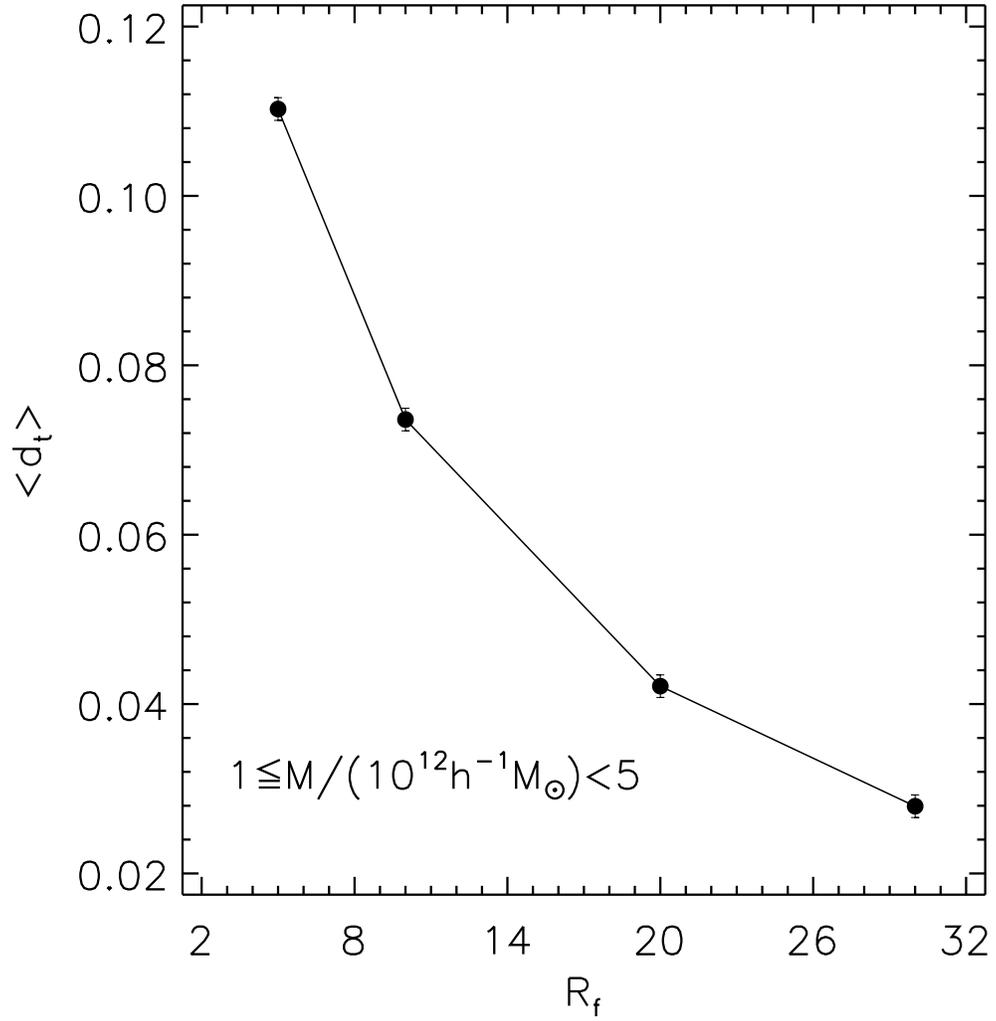}
\caption{Mean values of the shape correlation parameter $d_{t}$ averaged over the high-mass galactic halos 
as a function of the smoothing scale, $R_{f}$.}
\label{fig:dt_f}
\end{center}
\end{figure}

To numerically test the three analytic models, we have utilized the density fields and the Rockstar halo 
catalogs extracted from the SMDPL simulations \citep{smdpl}. Constructing the unit traceless tidal tensor, $\hat{\bf T}$, smoothed on the 
scale of $R_{f}=5\,h^{-1}$Mpc from the density fields given on the $512^{3}$ grids that constitute the simulation box of volume 
$400^{3}\,h^{-3}\,{\rm Mpc}^{3}$ and selecting the galactic halos in the mass range of $0.5\le M/(10^{11}\,h^{-1}\,M_{\odot})\le 50$ 
from the Rockstar catalog, we have first numerically obtained the probability density functions of the tidally induced shape 
alignments, $\{p(\vert\hat{\bf u}_{i}\cdot\hat{\bf e}\vert)\}_{i=1}^{3}$ (see Figures \ref{fig:pro_axis}-\ref{fig:pro_axis_filter}). 
The numerical results have clearly shown that $\hat{\bf e}$ has a tendency to be strongly aligned (anti-aligned) with 
$\hat{\bf u}_{3}$ ($\hat{\bf u}_{1}$) but  no correlation with $\hat{\bf u}_{2}$. 
Investigating the dependence of the strength of the tidally induced shape alignments on $M$, $R_{f}$, and the type of the cosmic 
web, it has been found that the more massive galactic halos yield stronger $\hat{\bf u}_{3}$-$\hat{\bf e}$ alignments 
($\hat{\bf u}_{1}$-$\hat{\bf e}$ anti-alignments) and that the increment of $R_{f}$ weakens the alignment tendency
(see Figures \ref{fig:dt_f}). 
These numerical results are consistent with what the previous works already found 
\citep{joa-etal13,zha-etal13,che-etal16,hil-etal17,xia-etal17,pir-etal18}. 

\begin{figure}
\begin{center}
\includegraphics[scale=1.0]{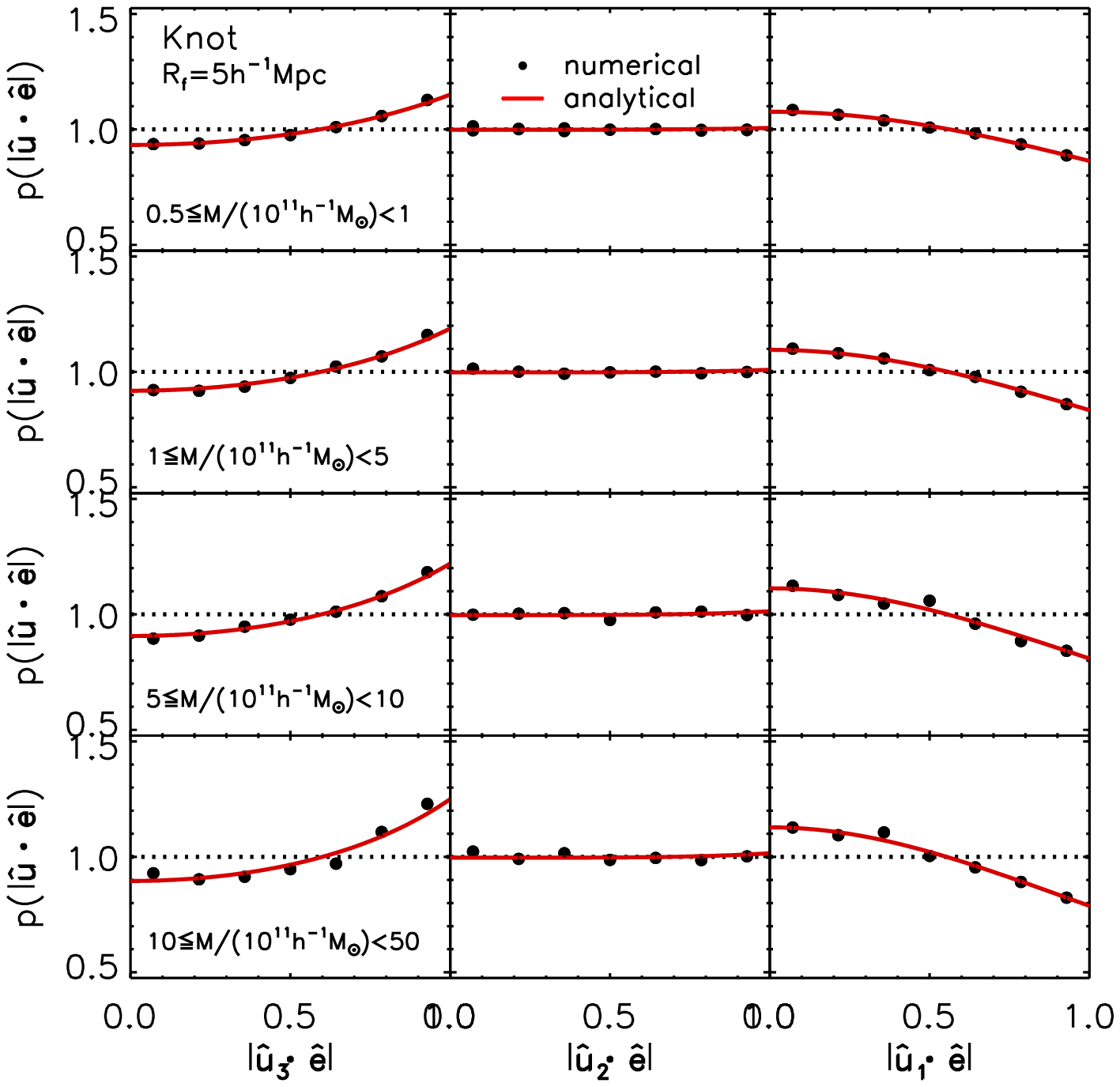}
\caption{Same as Figure \ref{fig:pro_axis} but with only those galactic halos located in the knot 
environments where all three eigenvalues of the tidal fields, $\lambda_{1},\ \lambda_{2},\ \lambda_{3}$,  
are positive.}
\label{fig:pro_axis_knot}
\end{center}
\end{figure}

\begin{figure}
\begin{center}
\includegraphics[scale=1.0]{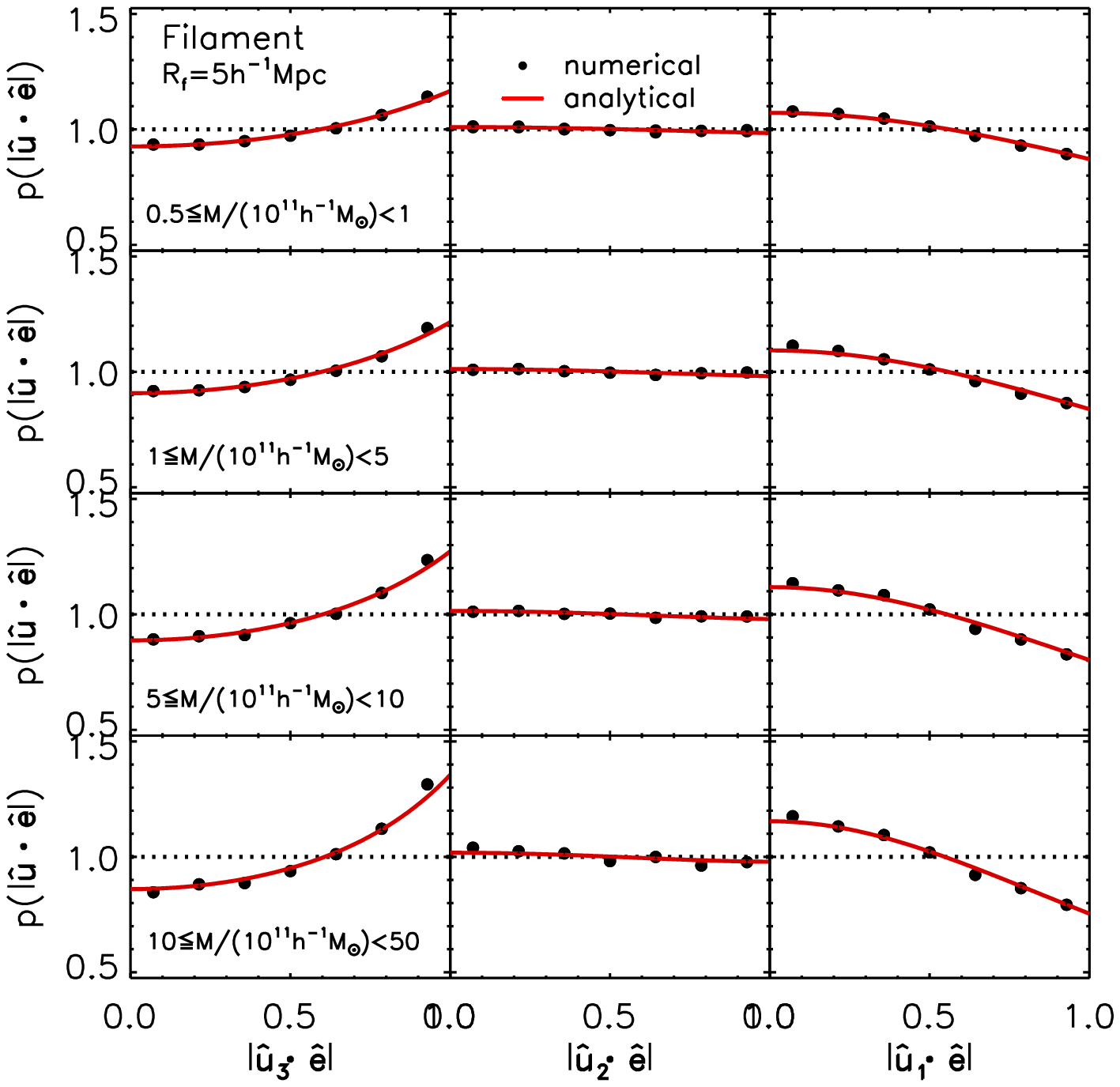}
\caption{Same as Figure \ref{fig:pro_axis} but with only those galactic halos located in the filament 
environments where $\lambda_{1}>\lambda_{2}>0>\lambda_{3}$.} 
\label{fig:pro_axis_fil}
\end{center}
\end{figure}

\begin{figure}
\begin{center}
\includegraphics[scale=1.0]{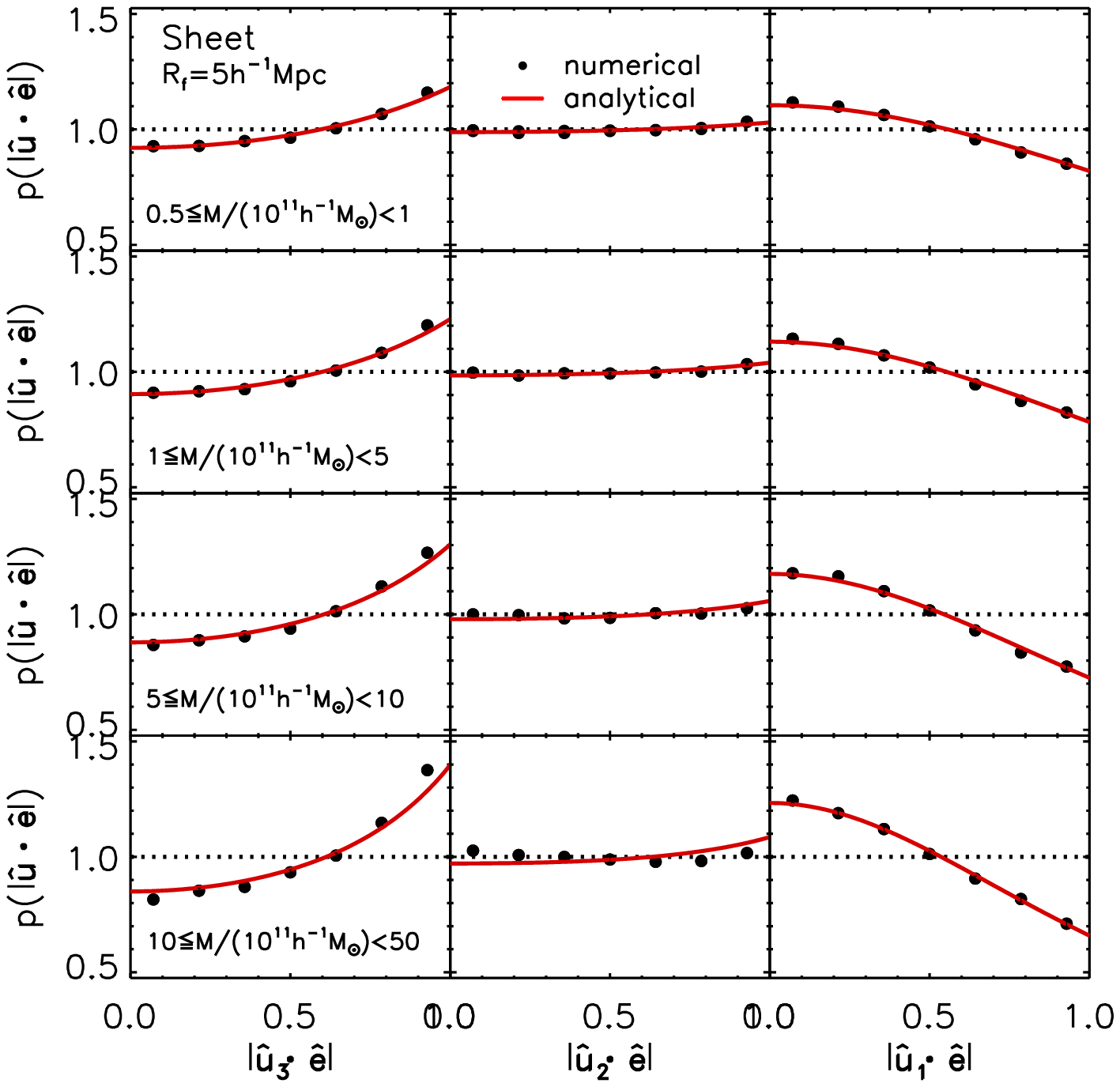}
\caption{Same as Figure \ref{fig:pro_axis} but with only those galactic halos located in the sheet
environments where $\lambda_{1}>0>\lambda_{2}>\lambda_{3}$.}
\label{fig:pro_axis_sheet}
\end{center}
\end{figure}

\begin{figure}
\begin{center}
\includegraphics[scale=1.0]{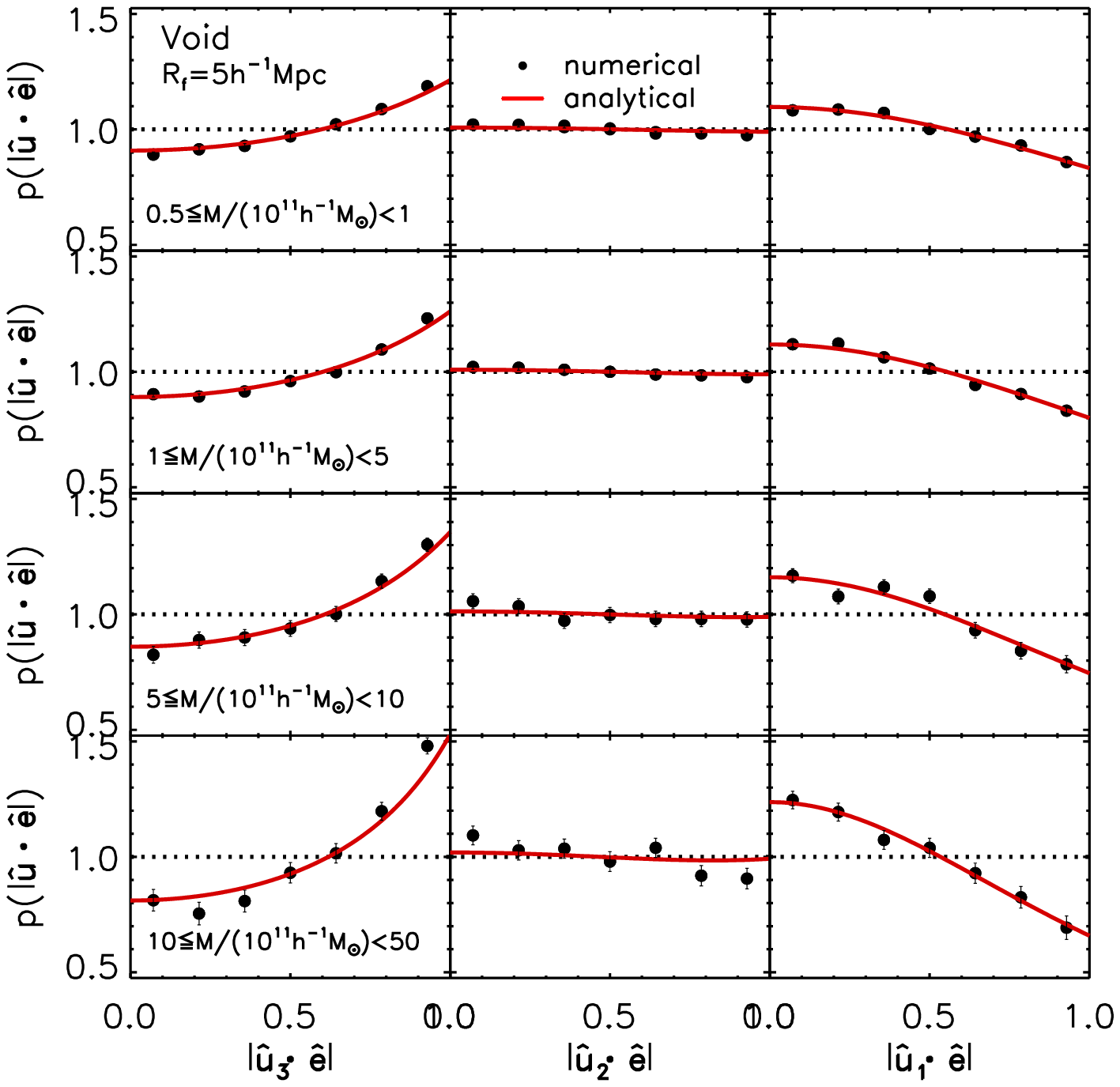}
\caption{Same as Figure \ref{fig:pro_axis} but with only those galactic halos located in the void 
environments where $0>\lambda_{1}>\lambda_{2}>\lambda_{3}$.}
\label{fig:pro_axis_void}
\end{center}
\end{figure}

\begin{figure}
\begin{center}
\includegraphics[scale=1.0]{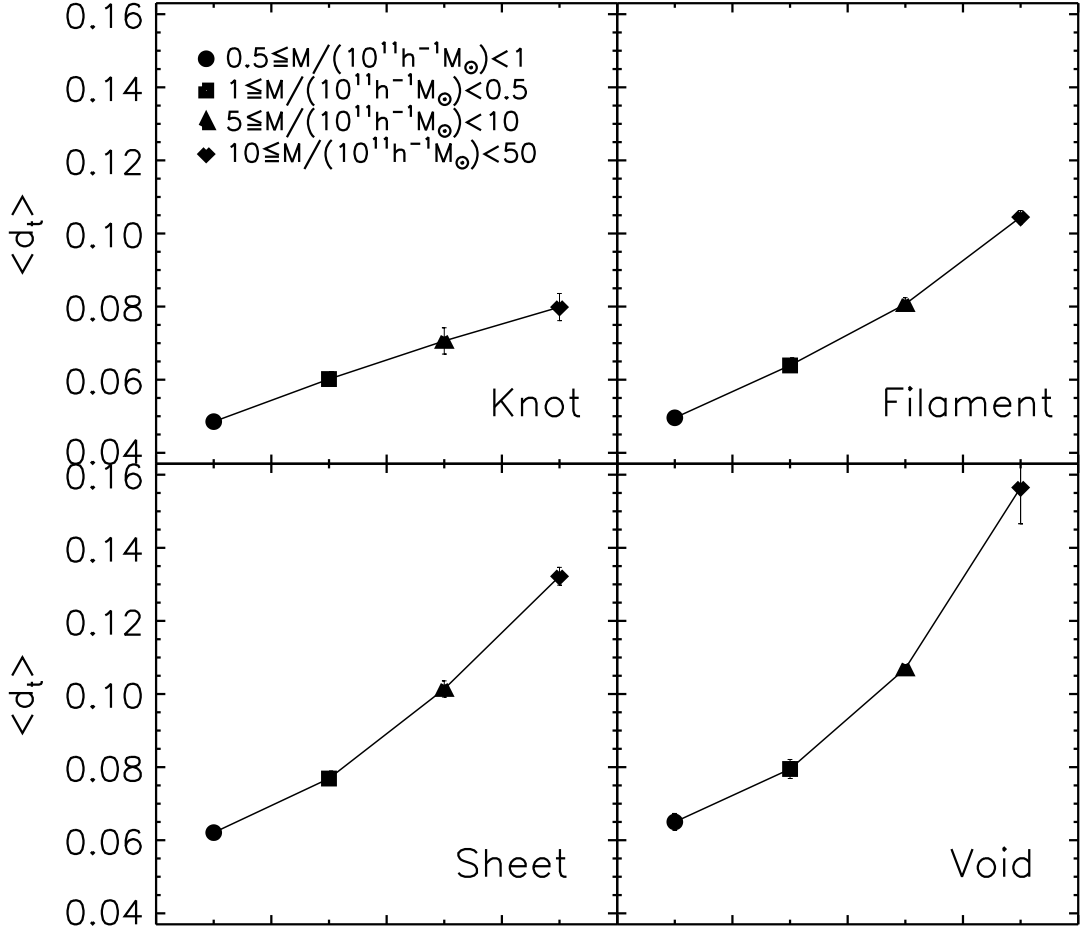}
\caption{Same as Figure \ref{fig:dt_m} but with those galactic halos embedded 
in four different types of the cosmic web.}
\label{fig:dt_web}
\end{center}
\end{figure}

The strongest (weakest) $\hat{\bf u}_{3}$-$\hat{\bf e}$ alignments are found from the void (knot) galactic halos 
(see Figures \ref{fig:pro_axis_knot}-\ref{fig:pro_axis_void}), which seem inconsistent with the previous numerical result 
that the DM halos showed the strongest shape alignments in the knot environments \citep{xia-etal17}. This inconsistency 
has been ascribed to the different classification schemes used in the two analyses. 
The sheet galactic halos yield much stronger shape alignment tendency than the knot and filament galactic halos in the whole  
mass range, which result is consistent with what \citet{hah-etal07a} found.  
In the lowest and low mass range ($0.5\le M/[10^{11}\,h^{-1}\,M_{\odot}]<5)$, the knot and filament galactic halos show similar strengths 
of the shape alignments. In the medium-mass ($5\le M/[10^{11}\,h^{-1}\,M_{\odot}]<10$) and 
high-mass ($10\le M/[10^{11}\,h^{-1}\,M_{\odot}]<50$) ranges, the shape alignments of the filament galactic halos become stronger 
than the knot counterparts (Figure \ref{fig:dt_web}). These numerical results imply that the void and sheet galactic halos retain best 
the tidally induced shape alignments, while the evolution of the galactic halos in the dense environments 
like the knots and filaments has an effect of deviating the directions of their shapes from the tidally induced inclinations. 

\begin{figure}
\begin{center}
\includegraphics[scale=1.0]{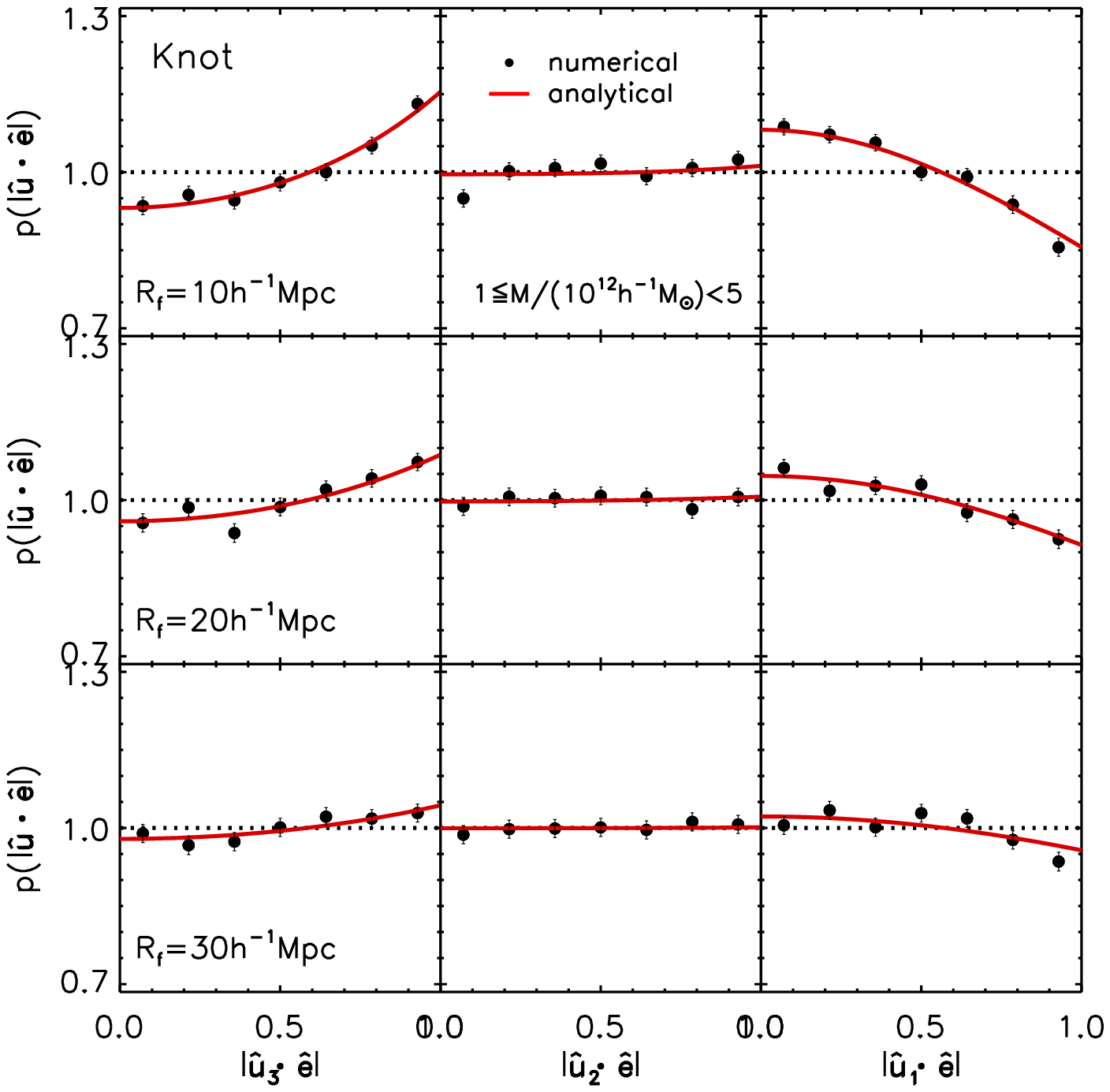}
\caption{Same as Figure \ref{fig:pro_axis_filter} but with only those galactic halos located in the knot 
environments.}
\label{fig:pro_axis_filter_knot}
\end{center}
\end{figure}

\begin{figure}
\begin{center}
\includegraphics[scale=1.0]{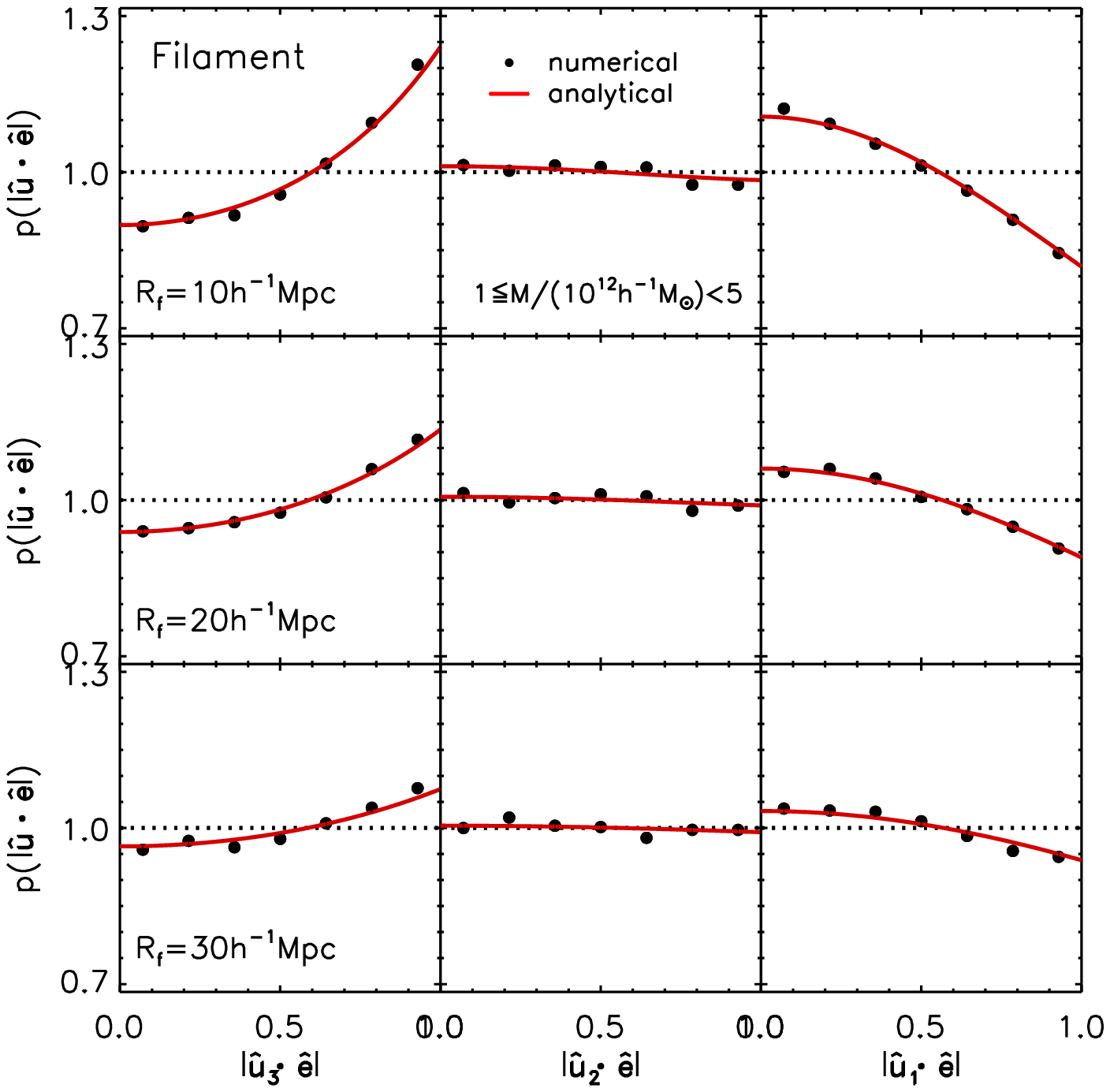}
\caption{Same as Figure \ref{fig:pro_axis_filter} but with only those galactic halos located in the filament 
environments where $\lambda_{1}>\lambda_{2}>0>\lambda_{3}$.}
\label{fig:pro_axis_filter_fil}
\end{center}
\end{figure}

\begin{figure}
\begin{center}
\includegraphics[scale=1.0]{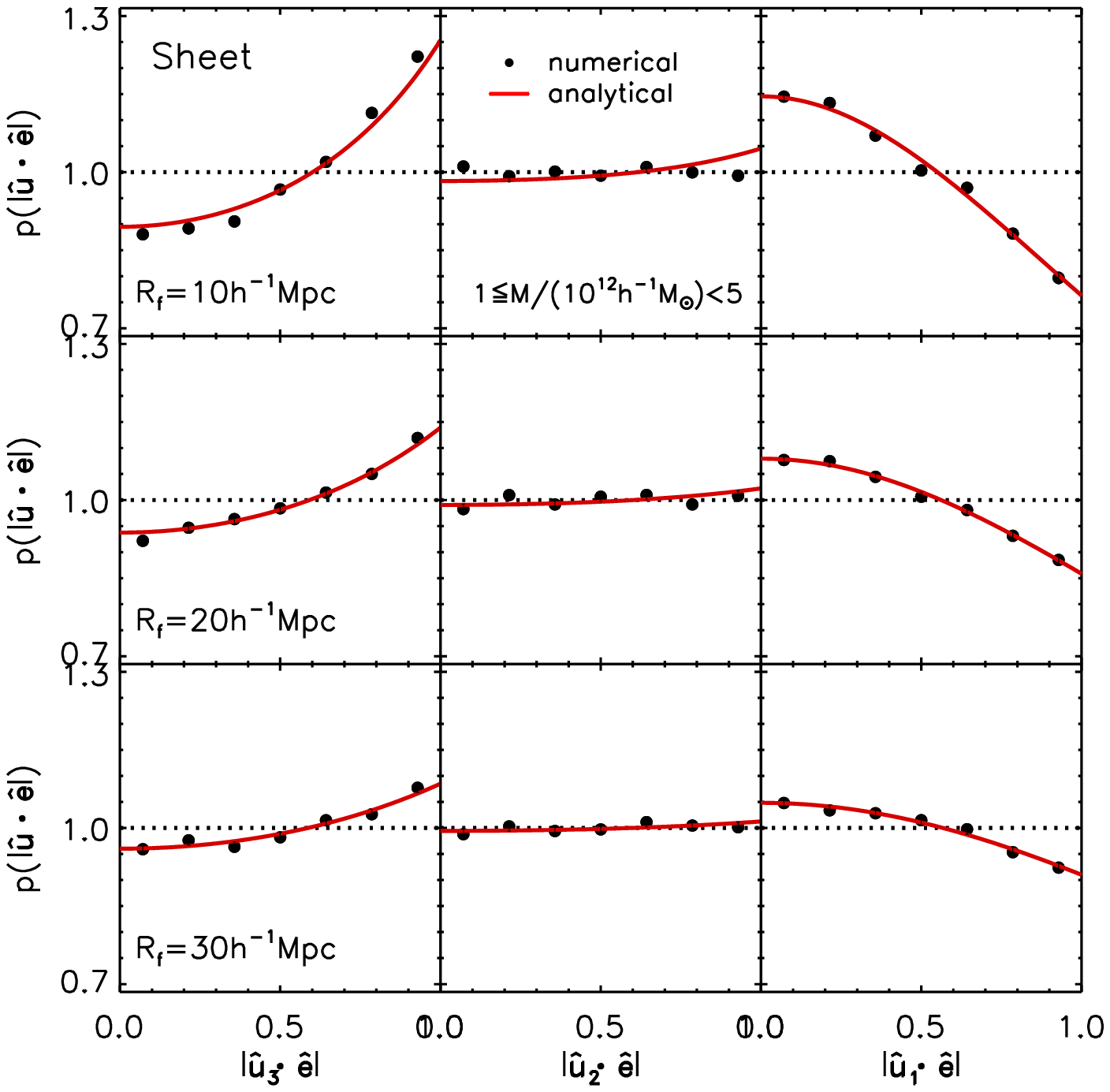}
\caption{Same as Figure \ref{fig:pro_axis_filter} but with only those galactic halos located in the sheet
environments where $\lambda_{1}>0>\lambda_{2}>\lambda_{3}$.}
\label{fig:pro_axis_filter_sheet}
\end{center}
\end{figure}

\begin{figure}
\begin{center}
\includegraphics[scale=1.0]{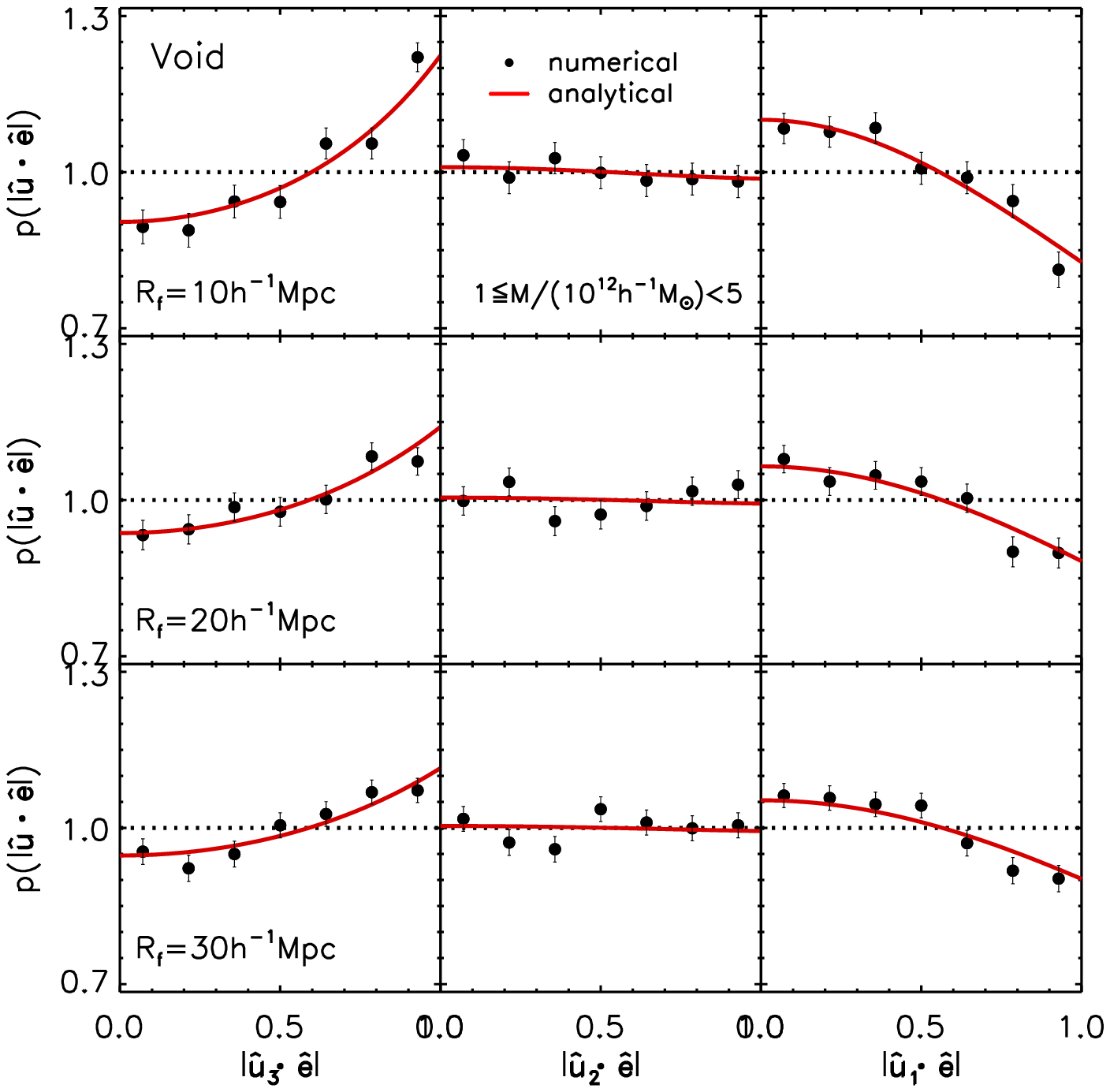}
\caption{Same as Figure \ref{fig:pro_axis_filter} but with only those galactic halos located in the  void
environments where $0>\lambda_{1}>\lambda_{2}>\lambda_{3}$.}
\label{fig:pro_axis_filter_void}
\end{center}
\end{figure}

\begin{figure}
\begin{center}
\includegraphics[scale=1.0]{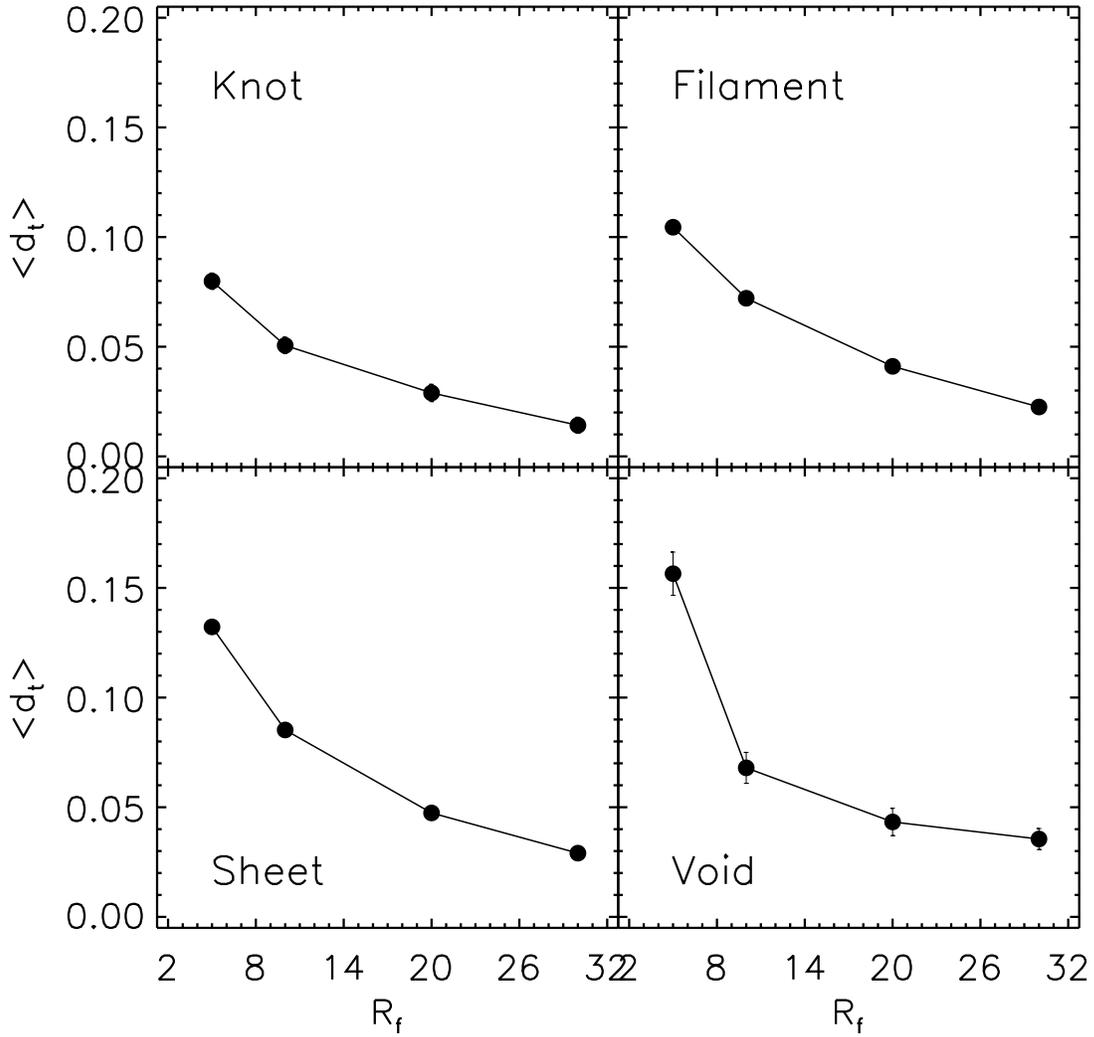}
\caption{Mean values of the shape correlation parameter averaged over the high-mass galactic halos 
belonging to the four different types of the cosmic web versus the smoothing scale $R_{f}$.}
\label{fig:dt_webf}
\end{center}
\end{figure}

The comparison with the numerical results revealed the success of the \texttt{model II} in  describing 
the amplitudes and behaviors of $\{p(\vert\hat{\bf u}_{i}\cdot\hat{\bf e}\vert)\}_{i=1}^{3}$, for all of the cases of 
$M$, $R_{f}$ and the type of the cosmic web. For the shape alignments, the \texttt{model III} turns out to be identical to 
the \texttt{model II}. For all of the four cases of the web type, the increment of $R_{f}$ has been found to 
decrease the strength of the tidally induced shape alignments but improve the agreements between the 
\texttt{model III} and the numerical results (Figure \ref{fig:pro_axis_filter_knot}-\ref{fig:dt_webf}). 
We interpret this result as an evidence supporting the scenario that the nonlinear evolution has an effect of 
diminishing the strength of the tidally induced shape alignments. 

\begin{figure}
\begin{center}
\includegraphics[scale=1.0]{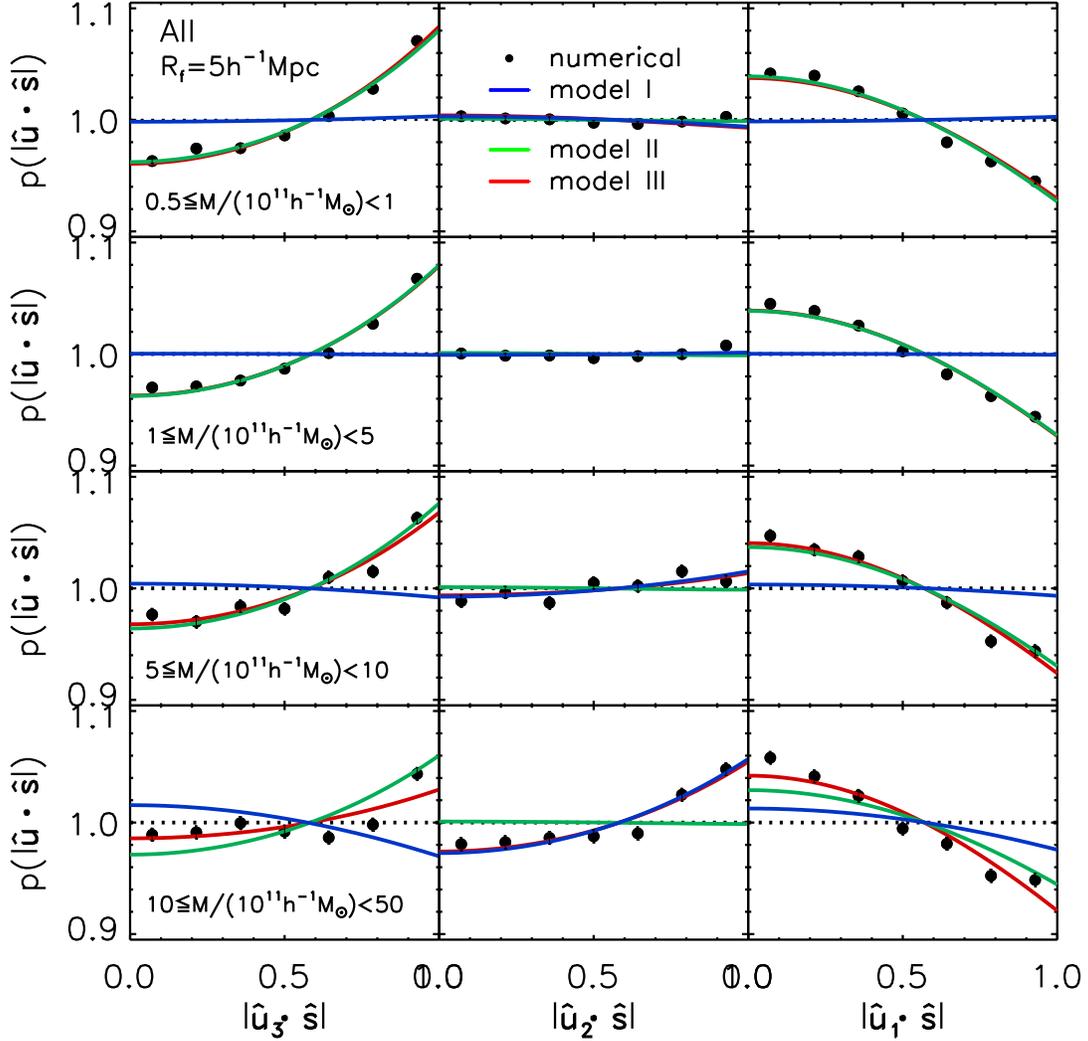}
\caption{Probability density distributions of three coordinates of the unit spin vectors, $\hat{\bf s}$, of 
the galactic halos in the principal frame spanned by three eigenvectors, $\{\hat{\bf u}_{1},\ \hat{\bf u}_{2},\ \hat{\bf u}_{3}\}$, 
of the local tidal fields smoothed on the scale of $R_{f}=5\,h^{-1}$Mpc, for four different ranges of the halo mass $M$. 
In each panel, the numerical results are plotted as black filled circular dots while the analytic 
model with the empirically determined parameters is shown as red solid line. The uniform probability density is 
depicted as black dotted line.}
\label{fig:pro_spin}
\end{center}
\end{figure}

\begin{figure}
\begin{center}
\includegraphics[scale=1.0]{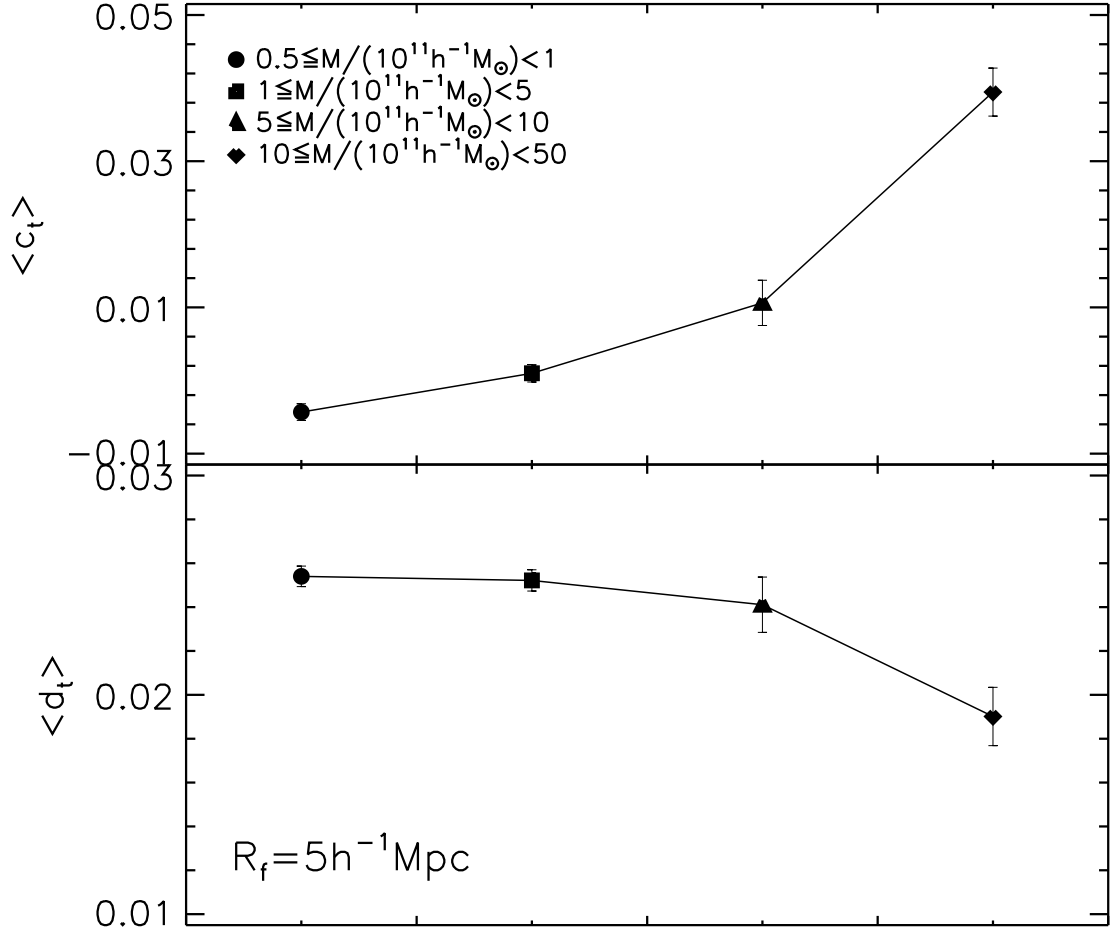}
\caption{Mean values of the first and second spin correlation parameters, $c_{t}$ and $d_{t}$, averaged over the galactic halos 
belonging to four different mass ranges for the case of $R_{f}=5\,h^{-1}$Mpc in the top and bottom panels, respectively.}
\label{fig:cdt_spin_m}
\end{center}
\end{figure}

In a similar manner, we have numerically determined the probability density functions of the tidally induced spin alignments, 
$\{p(\vert\hat{\bf u}_{i}\cdot\hat{\bf s}\vert)\}_{i=1}^{3}$,  explored their dependences on $M$, $R_{f}$ and the web type, 
and compared the results with the three analytic models. 
The tidally induced spin alignments have been found significant but quite weak compared with the shape alignments 
(Figures \ref{fig:pro_spin}-\ref{fig:cdt_spin_m}), consistent with the results from the previous works 
\citep[e.g.,][]{hah-etal07b,for-etal14,zha-etal15}.   
The occurrence of the spin-flip phenomenon has been witnessed. For the case of $R_{f}=5\,h^{-1}$Mpc, the lowest-mass, low-mass and 
medium-mass galactic halos show strong $\hat{\bf u}_{3}$-$\hat{\bf s}$ alignments and negligible $\hat{\bf u}_{2}$-$\hat{\bf s}$ 
alignments, while the high-mass galactic halos exhibit strong $\hat{\bf u}_{2}$-$\hat{\bf s}$ alignments, which results have confirmed 
the claims of the previous works \citep{ara-etal07,paz-etal08,zha-etal09,cod-etal12,lib-etal13,tro-etal13,dub-etal14,che-etal16,vee-etal18}. 

\begin{figure}
\begin{center}
\includegraphics[scale=1.0]{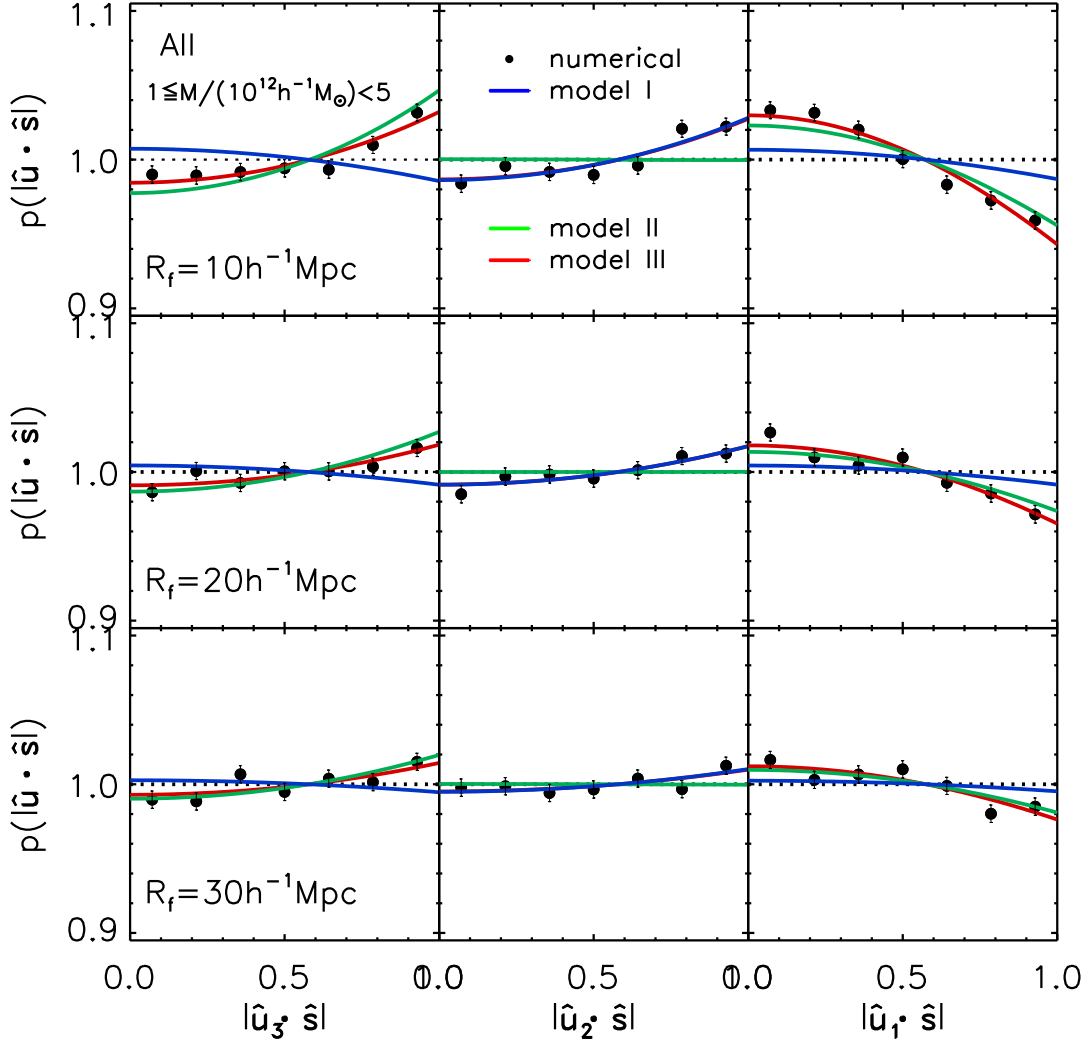}
\caption{Probability density distributions, 
$p(\vert\hat{\bf u}_{3}\cdot\hat{\bf e}\vert),\  p(\vert\hat{\bf u}_{2}\cdot\hat{\bf e}\vert),\ p(\vert\hat{\bf u}_{1}\cdot\hat{\bf e}\vert)$ 
of the high-mass galactic halos for three different cases of the smoothing scale $R_{f}$.}
\label{fig:pro_spin_filter}
\end{center}
\end{figure}

\begin{figure}
\begin{center}
\includegraphics[scale=1.0]{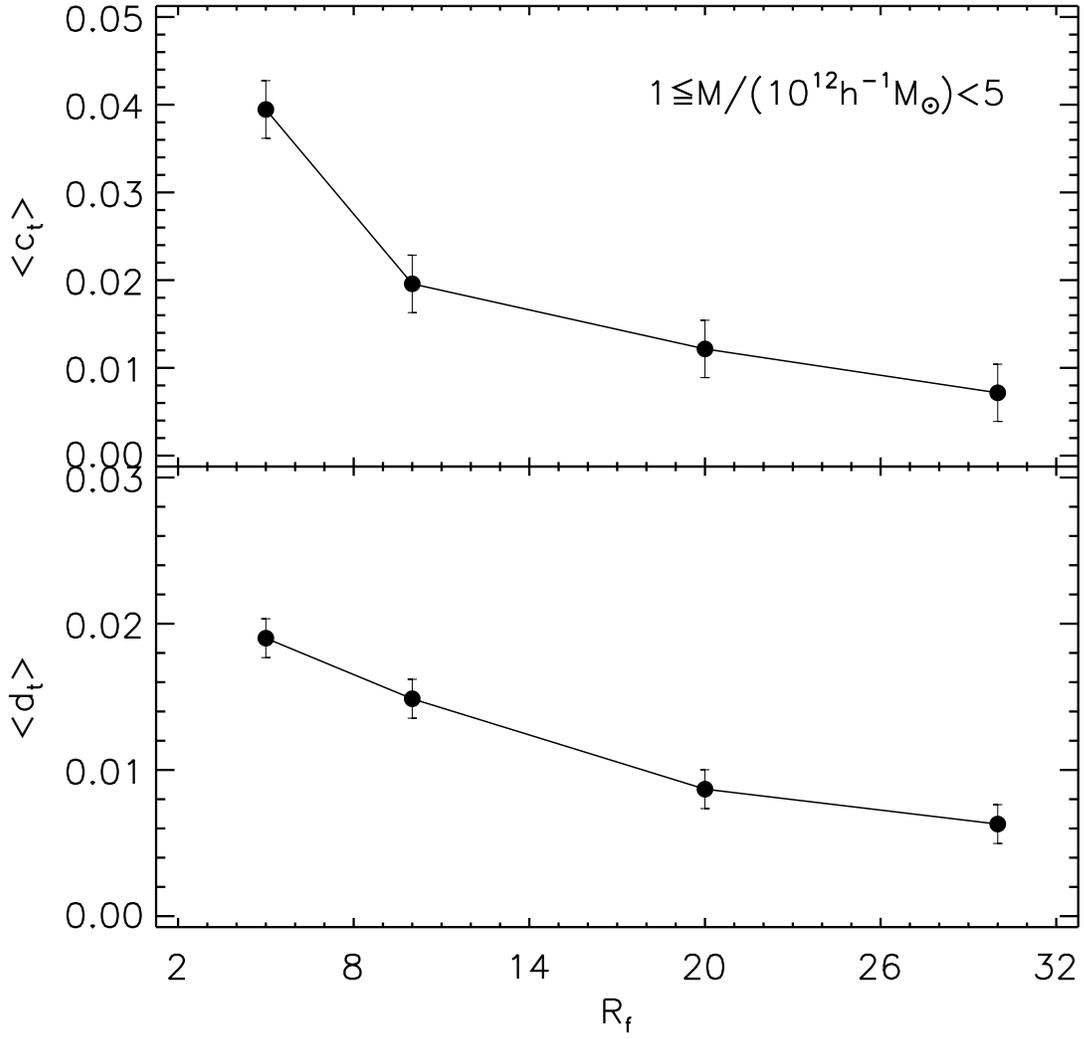}
\caption{Mean values of the first and second spin correlation parameters, $c_{t}$ and $d_{t}$, averaged over the 
high-mass galactic halos as a function of $R_{f}$ in the top and bottom panels, respectively.}
\label{fig:cdt_spin_f}
\end{center}
\end{figure}

However,  we have noted that the spin-flip does not occur abruptly at a certain fixed transition mass scale. Rather it is a gradual 
transition of the spin alignment tendency that proceeds over a broader mass range, depending on $R_{f}$ 
(Figures \ref{fig:pro_spin_filter}-\ref{fig:cdt_spin_f}). 
For the case of $R_{f}=5\,h^{-1}$Mpc, the high-mass galactic halos have been found to yield stronger 
$\hat{\bf u}_{2}$-$\hat{\bf s}$ and weaker but significant $\hat{\bf u}_{3}$-$\hat{\bf s}$ alignments, 
while the medium-mass galactic halos exhibit strong $\hat{\bf u}_{3}$-$\hat{\bf s}$ and much weaker 
$\hat{\bf u}_{2}$-$\hat{\bf s}$ alignments.  For the case of $R_{f}\ge 10\,h^{-1}$Mpc, however, the high-mass galactic halos 
exhibit stronger $\hat{\bf u}_{3}$-$\hat{\bf s}$ and weaker but significant $\hat{\bf u}_{2}$-$\hat{\bf s}$ alignments. 

\begin{figure}
\begin{center}
\includegraphics[scale=1.0]{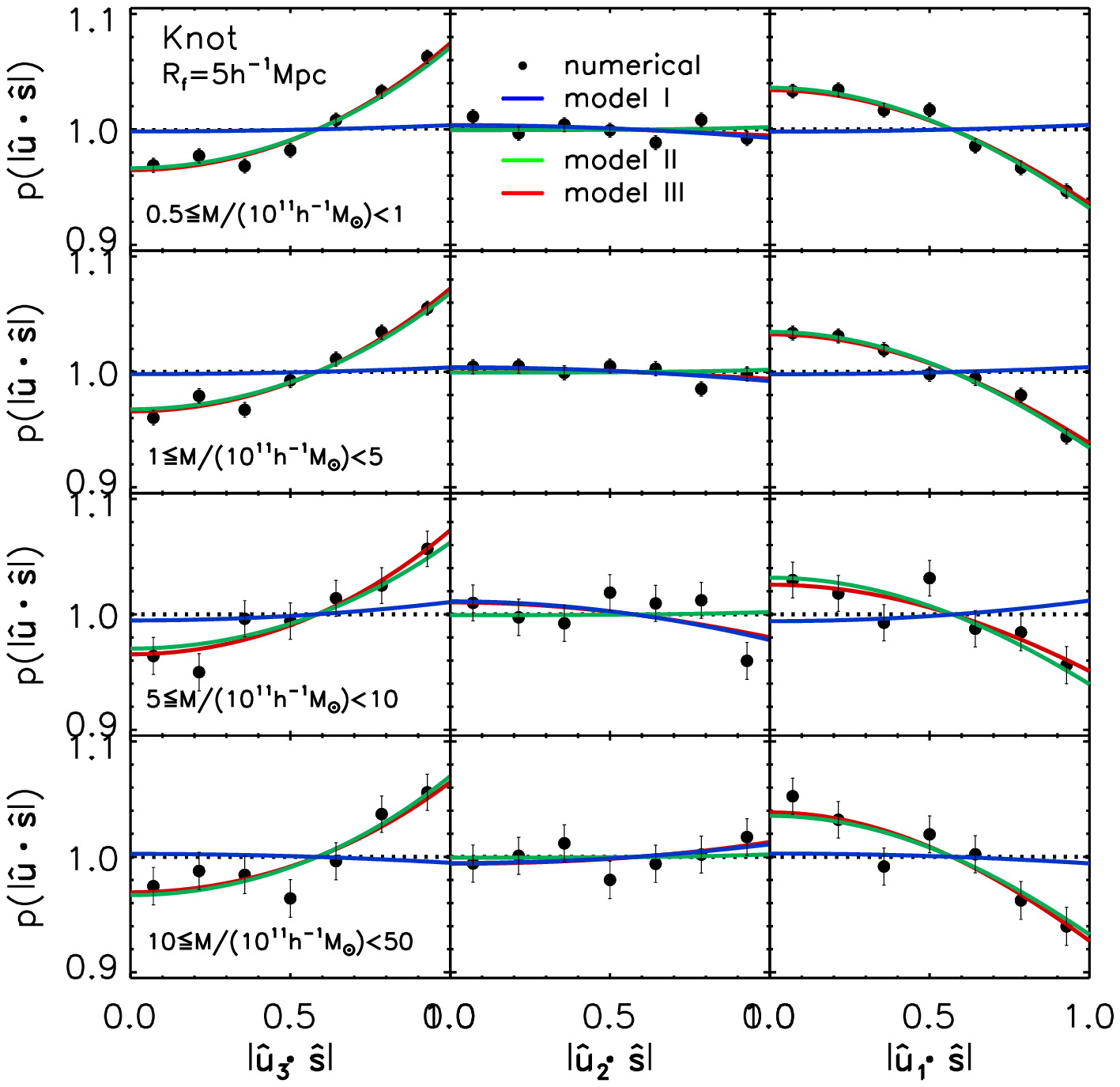}
\caption{Same as Figure \ref{fig:pro_spin} but with only those galactic halos located in the knot 
environments where all three eigenvalues of the local tidal fields, $\lambda_{1},\ \lambda_{2},\ \lambda_{3}$,  
are positive.}
\label{fig:pro_spin_knot}
\end{center}
\end{figure}

\begin{figure}
\begin{center}
\includegraphics[scale=1.0]{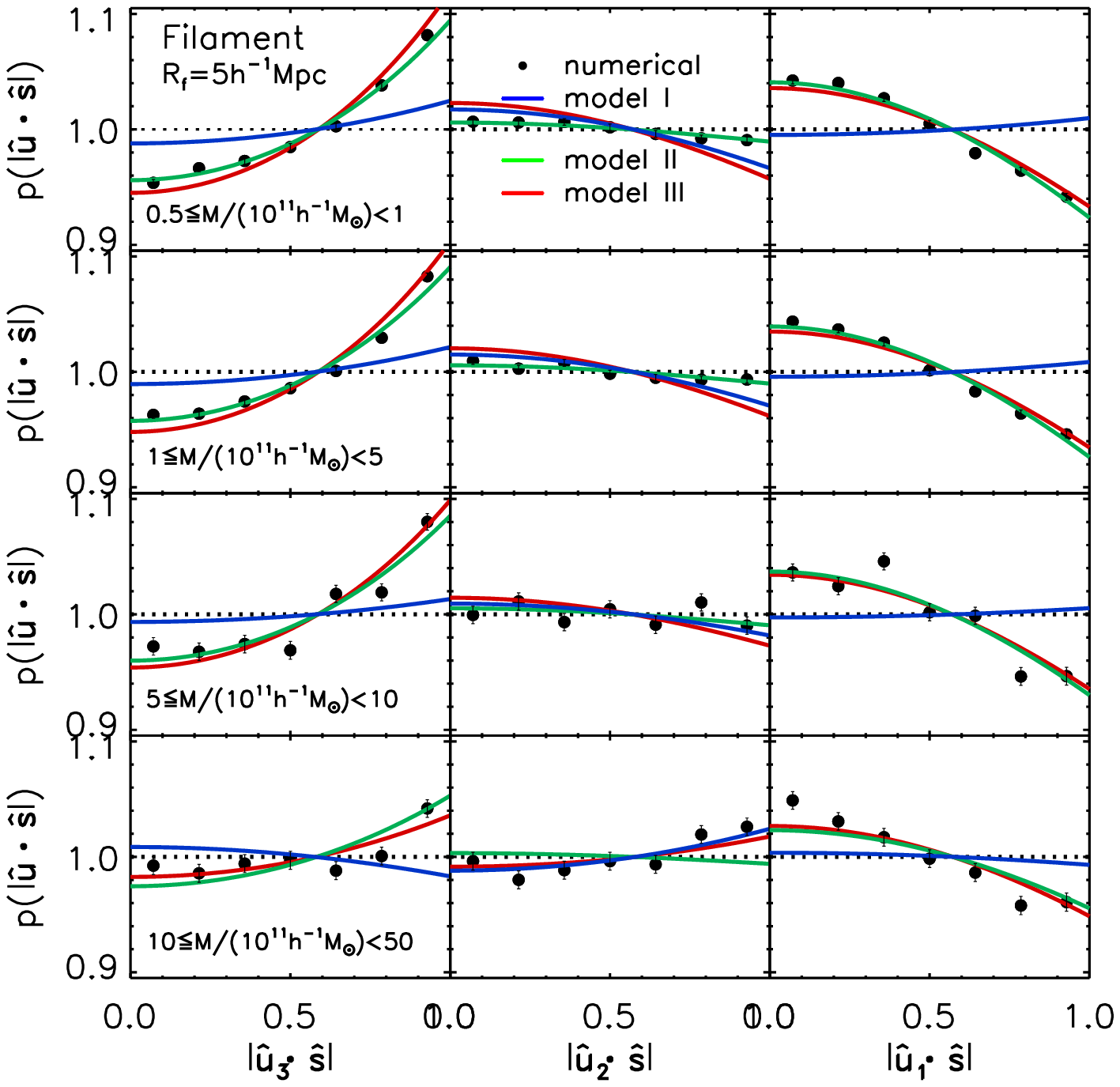}
\caption{Same as Figure \ref{fig:pro_spin} but with only those galactic halos located in the filament 
environments where $\lambda_{1}>\lambda_{2}>0>\lambda_{3}$.} 
\label{fig:pro_spin_fil}
\end{center}
\end{figure}

\begin{figure}
\begin{center}
\includegraphics[scale=1.0]{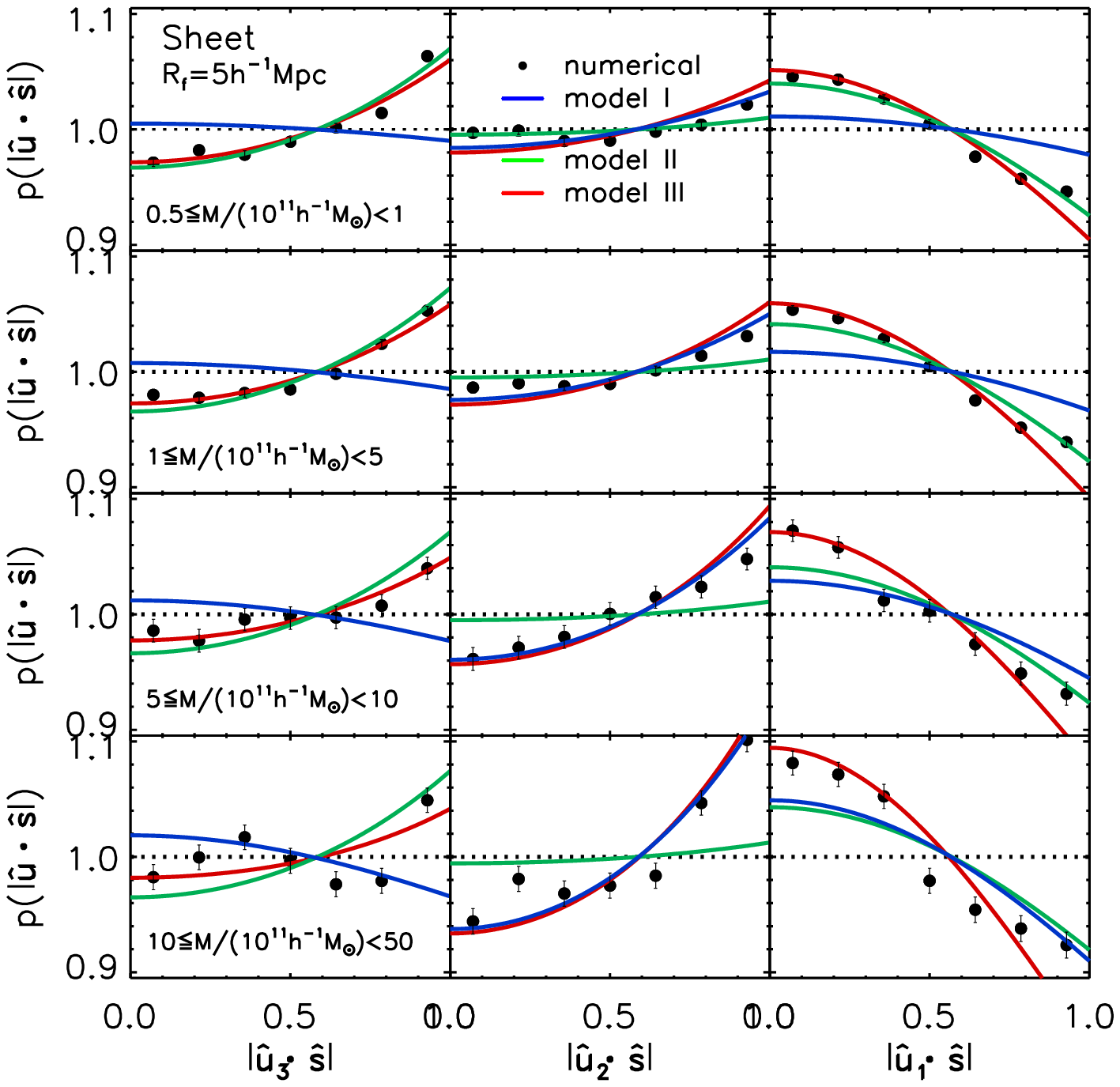}
\caption{Same as Figure \ref{fig:pro_spin} but with only those galactic halos located in the sheet
environments where $\lambda_{1}>0>\lambda_{2}>\lambda_{3}$.}
\label{fig:pro_spin_sheet}
\end{center}
\end{figure}

\begin{figure}
\begin{center}
\includegraphics[scale=1.0]{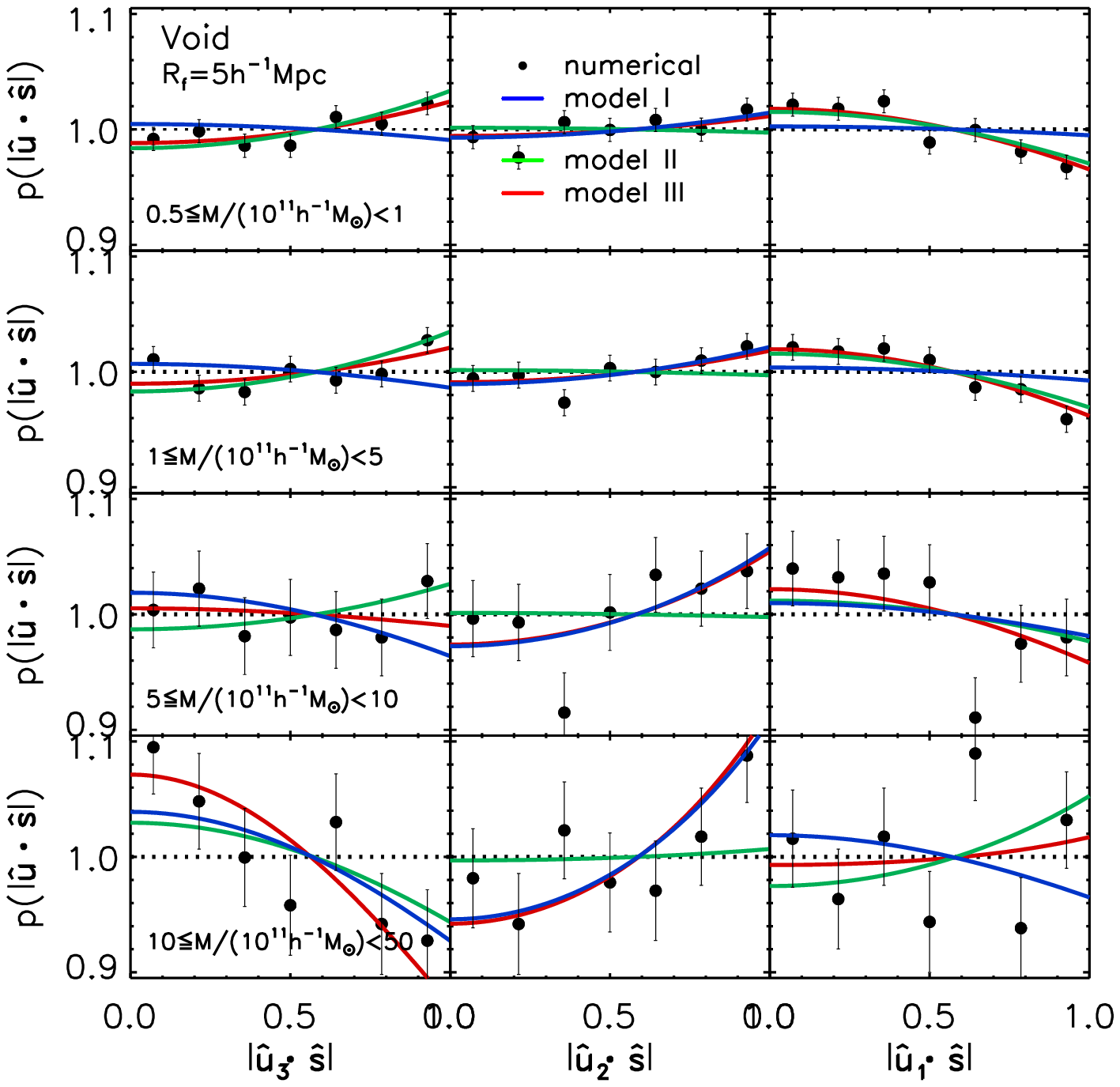}
\caption{Same as Figure \ref{fig:pro_spin} but with only those galactic halos located in the void 
environments where $0>\lambda_{1}>\lambda_{2}>\lambda_{3}$.}
\label{fig:pro_spin_void}
\end{center}
\end{figure}

\begin{figure}
\begin{center}
\includegraphics[scale=1.0]{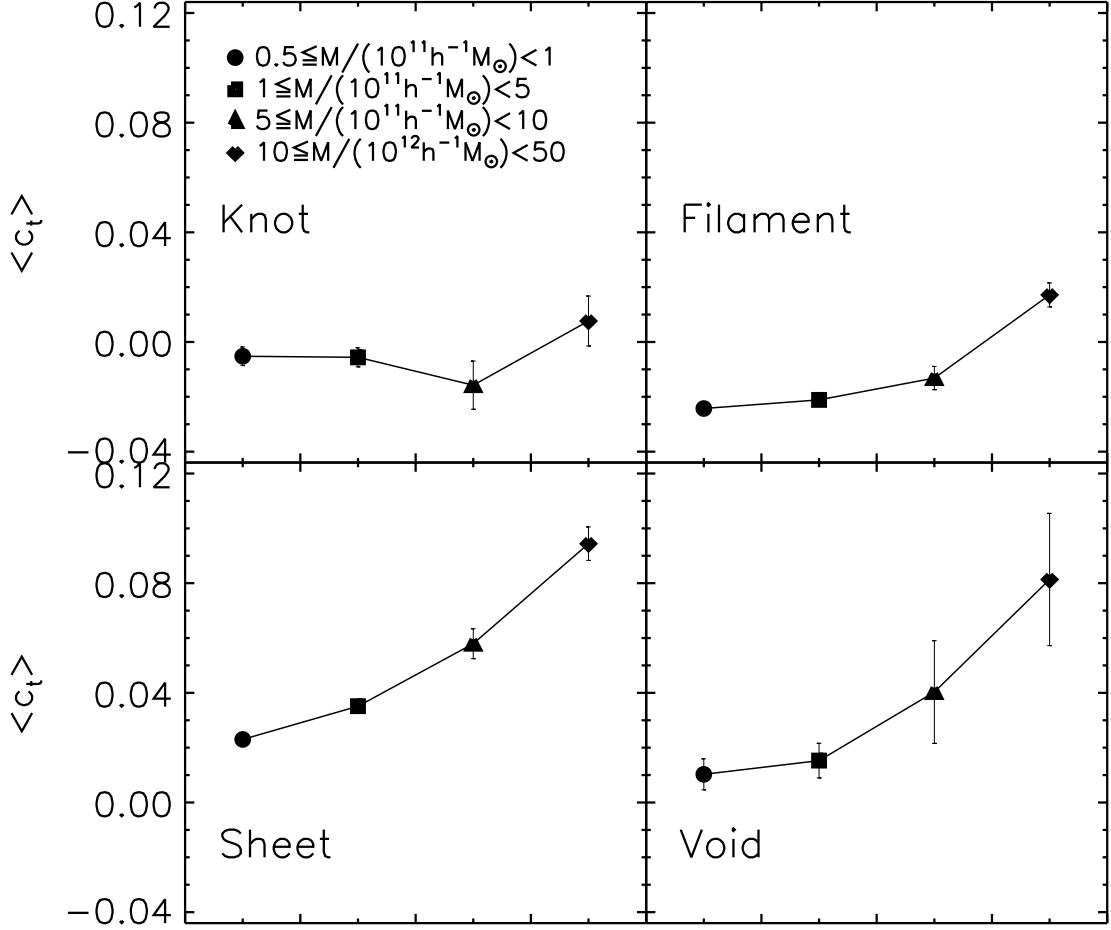}
\caption{Mean values of the first spin correlation parameter $c_{t}$ averaged over the knot (top-left panel), 
filament (top-right panel), sheet (bottom-left panel), and void (bottom-right panel) galactic halos in four different 
mass ranges. The smoothing scale $R_{f}$ is set at $5\,h^{-1}$Mpc.}
\label{fig:ct_spin_web}
\end{center}
\end{figure}

\begin{figure}
\begin{center}
\includegraphics[scale=1.0]{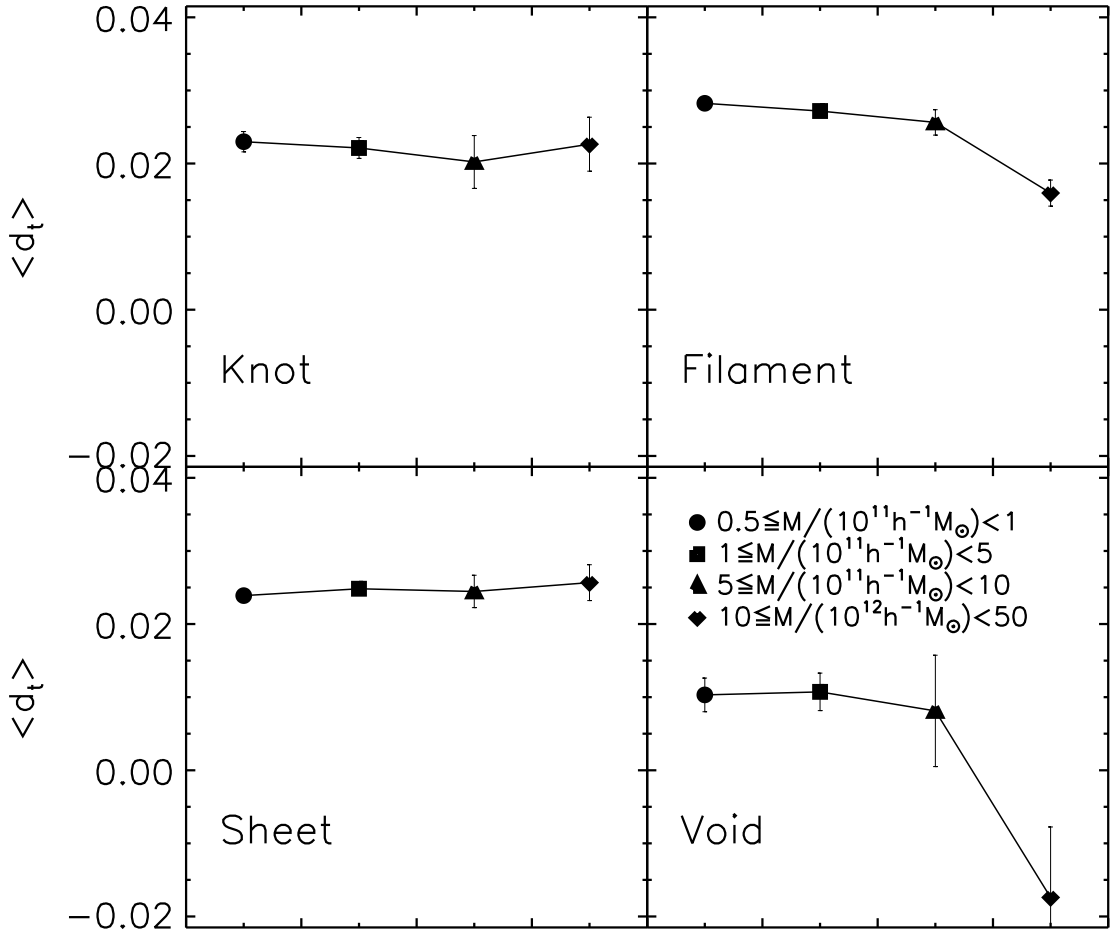}
\caption{Same as Figure \ref{fig:ct_spin_web} but for the second spin correlation parameter $d_{t}$.}
\label{fig:dt_spin_web}
\end{center}
\end{figure}

\begin{figure}
\begin{center}
\includegraphics[scale=1.0]{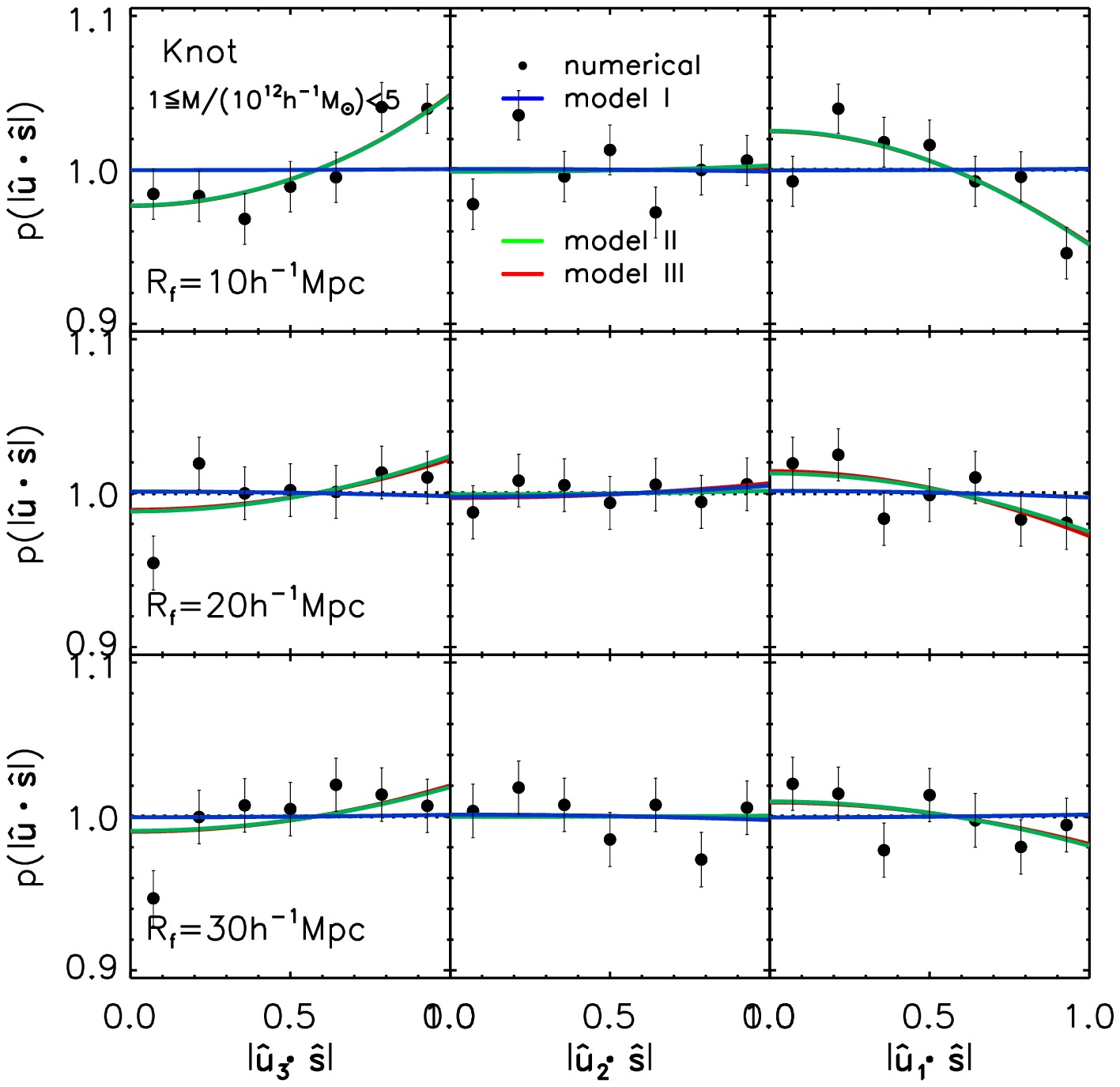}
\caption{Same as Figure \ref{fig:pro_spin_filter} but with only those galactic halos located in the knot 
environments.}
\label{fig:pro_spin_filter_knot}
\end{center}
\end{figure}

\begin{figure}
\begin{center}
\includegraphics[scale=1.0]{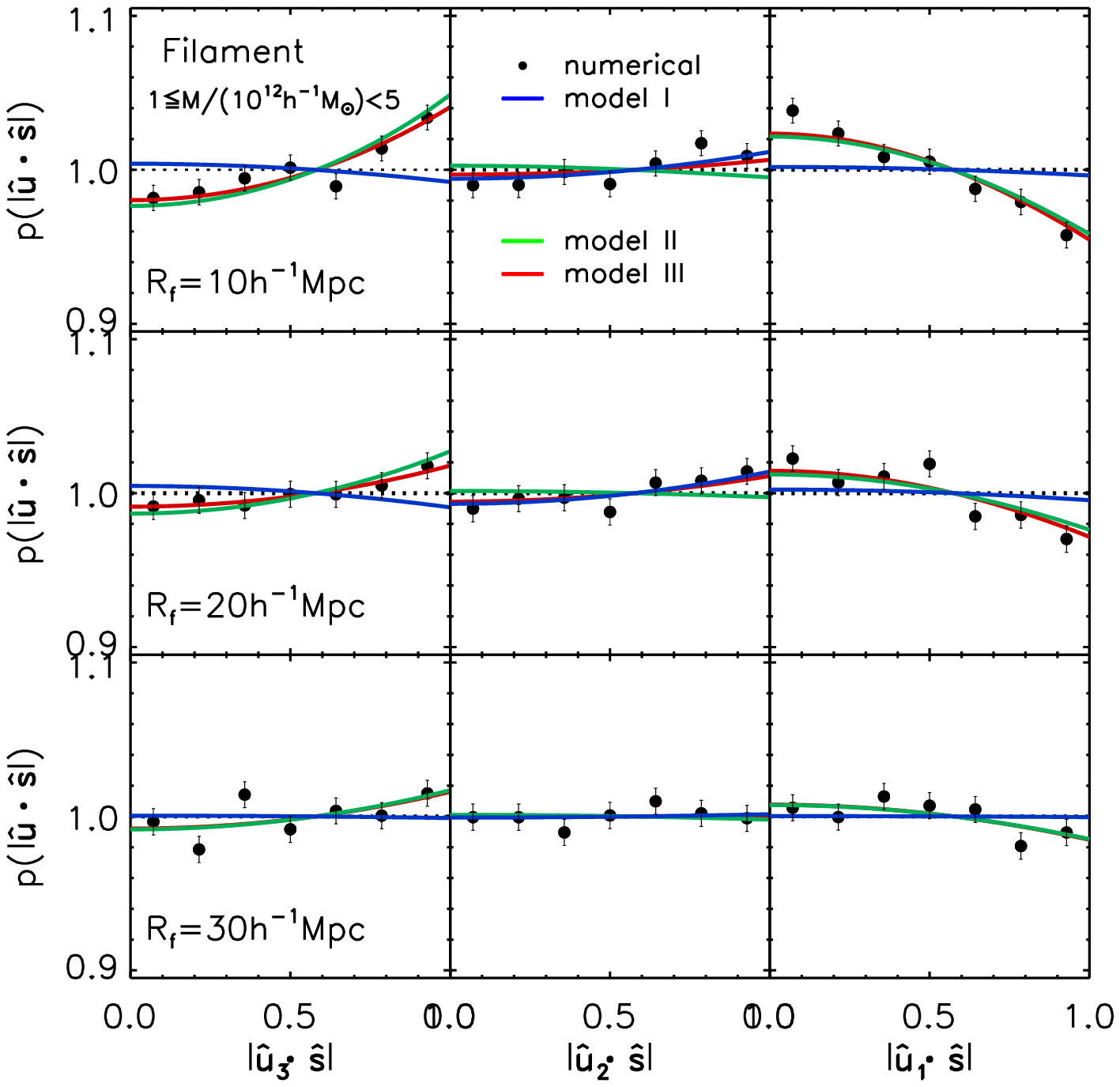}
\caption{Same as Figure \ref{fig:pro_spin_filter} but with only those galactic halos located in the filament 
environments where $\lambda_{1}>\lambda_{2}>0>\lambda_{3}$.}
\label{fig:pro_spin_filter_fil}
\end{center}
\end{figure}

\begin{figure}
\begin{center}
\includegraphics[scale=1.0]{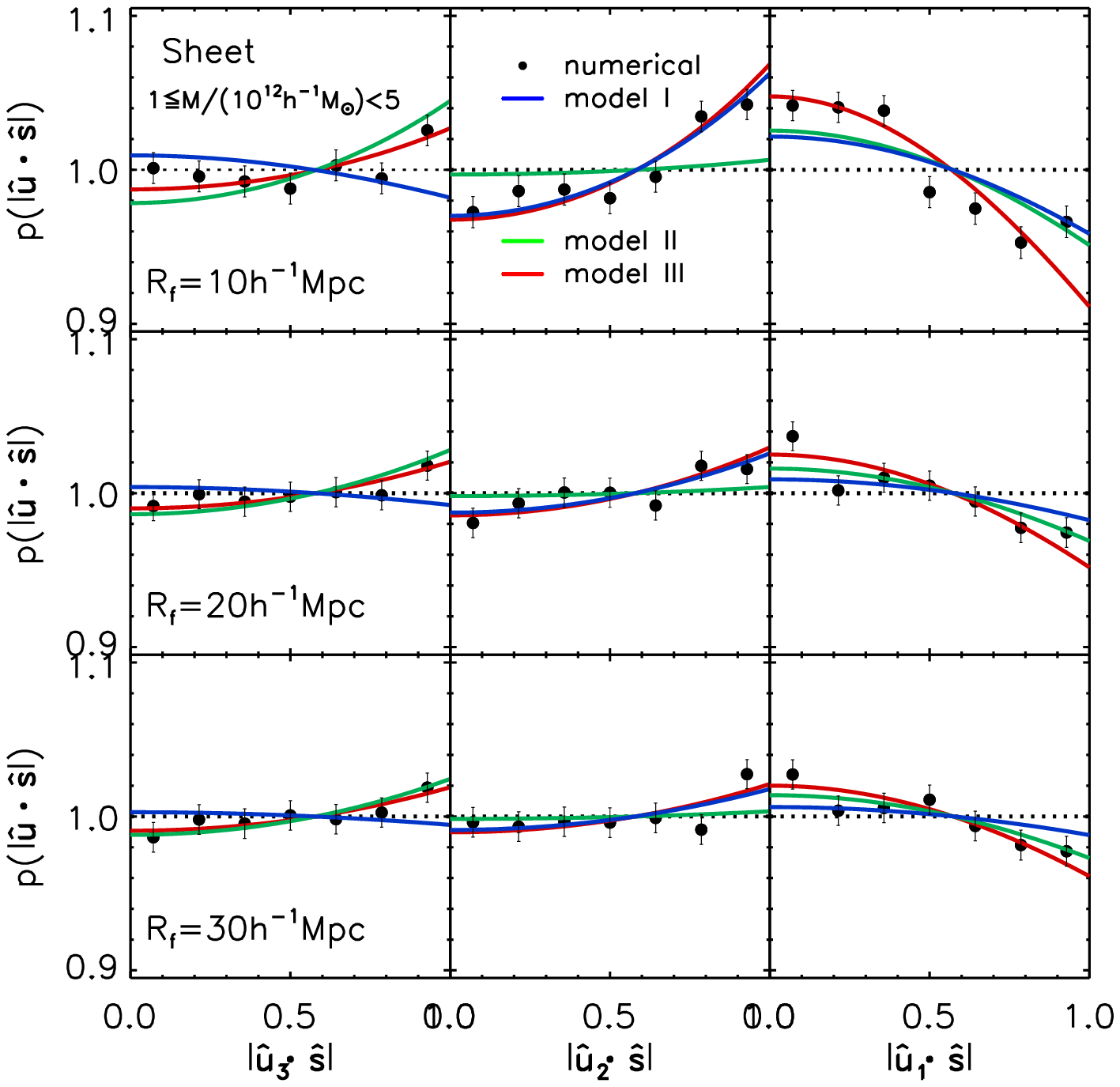}
\caption{Same as Figure \ref{fig:pro_spin_filter} but with only those galactic halos located in the sheet
environments where $\lambda_{1}>0>\lambda_{2}>\lambda_{3}$.}
\label{fig:pro_spin_filter_sheet}
\end{center}
\end{figure}

\begin{figure}
\begin{center}
\includegraphics[scale=1.0]{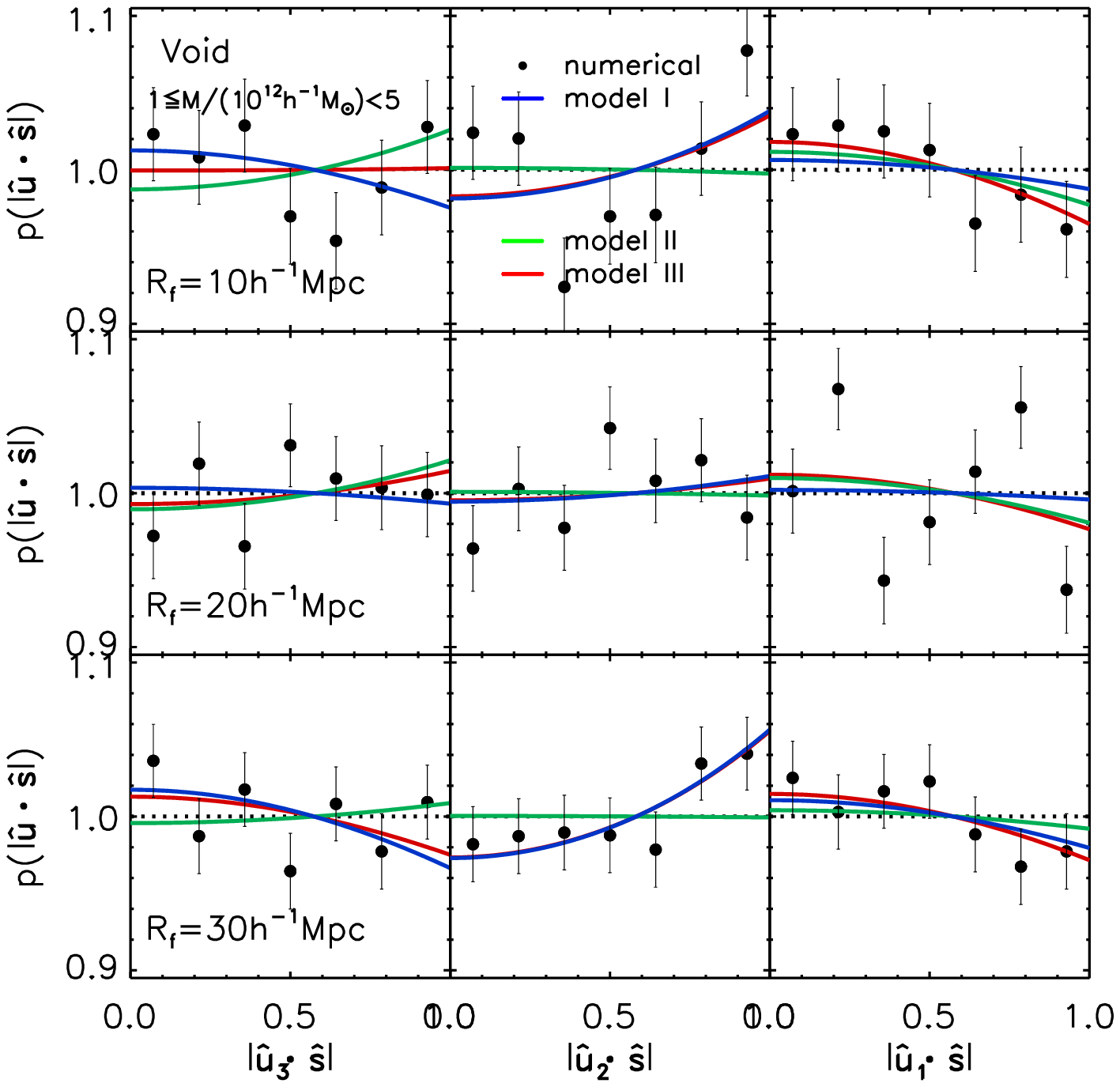}
\caption{Same as Figure \ref{fig:pro_spin_filter} but with only those galactic halos located in the void 
environments where $0>\lambda_{1}>\lambda_{2}>\lambda_{3}$.}
\label{fig:pro_spin_filter_void}
\end{center}
\end{figure}

The strengths of the tidally induced spin alignments have been also found to sensitively vary with the types of the cosmic web 
(see Figures \ref{fig:pro_spin_knot}-\ref{fig:pro_spin_filter_void}), which supports the claim of \citet{lib-etal13}.  
The strongest (weakest) signals of the tidally induced spin alignments have been found from the sheet (void) galactic halos, while the 
filament galactic halos have been found to have stronger spin alignments than the knot counterparts in the whole mass range 
( Figures \ref{fig:ct_spin_web}-\ref{fig:dt_spin_web}). These results are inconsistent with the observational finding of \citet{zha-etal15} 
that the knot galaxies exhibited the strongest signals of the spin alignments. We have suspected that this inconsistency might be 
related to  the construction of the tidal field from the galaxy groups and the determination of the spin axes of the galaxies from their 
stellar components in the observational analysis. 

Determining empirically $\langle c_{t}\rangle$ and $\langle d_{t}\rangle$ from the numerical data 
(Figures \ref{fig:ct_spin_web}-\ref{fig:dt_spin_web}) and defining the condition for the occurrence of the spin flip as  
$\langle c_{t}\rangle > \langle d_{t}\rangle$, we have quantitatively investigated how the occurrence and the transition mass 
scale, $M_{t}$, of the spin-flip phenomenon depend on the size and type of the cosmic web and found the following: 
\begin{enumerate}
\item 
Regardless of the web type, the transition mass scale, $M_{t}$, of the spin-flip increases with the increment of $R_{f}$. 
\item 
The knot galactic halos do no show any spin-flip phenomenon. That is, the unit spin vectors, $\hat{\bf s}$, of the knot galactic halos 
are always preferentially aligned with $\hat{\bf u}_{3}$ rather than with $\hat{\bf u}_{2}$ in the whole mass range, regardless of the value 
of $R_{f}$ (Figure \ref{fig:pro_spin_filter_knot}). 
\item
For the case of the filament galactic halos, the spin flip occurs around $M_{\rm t}\sim 5\times 10^{12}\,h^{-1}\,M_{\odot}$ when 
$R_{f}=5\,h^{-1}$Mpc. At the larger scale of $R_{f}>5\,h^{-1}$Mpc, the value of $M_{\rm t}$ exceeds the galactic mass scales, i.e, 
$M_{\rm t}> 5\times 10^{12}\,h^{-1}\,M_{\odot}$ (Figure \ref{fig:pro_spin_filter_fil}).  
\item
In the sheet environment, the transition mass scale has a lower value than in the filaments: 
$M_{\rm t}\sim 10^{12}\,h^{-1}\,M_{\odot}$ when $R_{f}=5\,h^{-1}$Mpc. Only when $R_{f}$ reaches $30\,h^{-1}$Mpc, 
the value of $M_{\rm t}$ becomes larger than the galactic mass scale (Figure \ref{fig:pro_spin_filter_sheet}). 
\item
The void galactic halos yield the lowest transition mass scale, $M_{\rm t}\sim 5\times 10^{11}\,h^{-1}\,M_{\odot}$ when 
$R_{f}=5\,h^{-1}$Mpc. At the larger scales, the number of the void galactic halos is too low to produce any significant signals 
(Figure \ref{fig:pro_spin_filter_void}). 
\end{enumerate}

It is interesting to note that our results on the web and mass dependence of the spin-flip phenomenon are consistent with the theoretical 
explanation of \citet{cod-etal15}, according to which the misalignments between the inertia momentum and tidal tensors in the anisotropic 
environments like the filaments and sheets are largely responsible for the occurrence of the spin flip. In line with their theoretical 
explanation, we interpret no occurrence of the spin flip in the knot environments as an evidence for the stronger alignments 
between the two tensors in the dense environments. In other words, in the knot regions where the tidal tensors are 
more isotropic, the inertia momentum and tidal tensors may be more strongly aligned with each other, which plays a role in 
suppressing the occurrence of the spin-flip of the knot galaxies. 

It has also been clearly demonstrated in the current work that the \texttt{model III} succeeds in describing consistently and simultaneously 
the numerical results of the tidally induced shape and spin alignments for all of the cases of $M$, $R_{f}$ and type of the cosmic web, 
while the \texttt{model I} and \texttt{model II} fail.  Showing that the \texttt{model III} works better as $R_{f}$ increases, we have 
ascribed the slight mismatches between the numerical results and the \texttt{model III} to the inaccuracies caused by the 
approximations of $p({\bf s}\vert{\bf T})$ as a multivariate Gaussian distribution and $\hat{\bf T}$ as a Gaussian random field made in 
the construction of the \texttt{model III}. We also suspect that the uncertainties in the measurements of $\hat{\bf s}$ and 
$\hat{\bf e}$ caused by the simple assumptions of each galactic halo having a perfect ellipsoidal shape and no substructure 
in a completely relaxed dynamical state must contribute to the mismatches. 

We conclude that the \texttt{model III} is an effective practical model for the spin and the shape alignments of the galactic halos 
with the large-scale tidal fields, providing an analytic tool with which the condition of the spin flip occurrence as well as its dependence on 
the properties of the large-scale structures can be quantitatively described. Its good accord with the numerical results supports the 
scenario that the occurrence of the spin flip phenomenon is associated more with the geometrical properties of the large-scale tidal field as 
well as the interactions of the galactic halos with the cosmic web rather than with the physical processes during the nonlinear evolution 
\citep[see][]{lib-etal13,cod-etal15,WK17,vee-etal18}. 

Given that the \texttt{model III} is expressed in terms of the linear quantities, it may provide another independent probe of the background 
cosmology.  For this purpose, however, a couple of back-up works will have to be done. First, as suspected in our analysis, differences in 
the schemes used to to construct the tidal fields, to measure the shape and spin axes of the galaxies, and to classify the cosmic web 
would yield different patterns in the dependence of the tidally induced shape and spin alignments on the sizes and types of the cosmic 
web. Thus, it will be necessary to test the robustness of the \texttt{model III} against the variations of the schemes. 
Second, it will be also essential to examine its validity using the numerical results for alternative cosmologies such as models with modified 
gravity, coupled dark energy, massive neutrinos, primordial non-Gaussianity, anisotropic inflation and so forth. 
Our future work is in this direction.

\acknowledgements

ÒThe CosmoSim database used in this paper is a service by the Leibniz-Institute for Astrophysics Potsdam (AIP).
The MultiDark database was developed in cooperation with the Spanish MultiDark Consolider Project CSD2009-00064.Ó
I gratefully acknowledge the Gauss Centre for Supercomputing e.V. (www.gauss-centre.eu) and the Partnership for Advanced 
Supercomputing in Europe (PRACE, www.prace-ri.eu) for funding the MultiDark simulation project by providing computing time on 
the GCS Supercomputer SuperMUC at Leibniz Supercomputing Centre (LRZ, www.lrz.de).
The Bolshoi simulations have been performed within the Bolshoi project of the University of California High-Performance 
AstroComputing Center (UC-HiPACC) and were run at the NASA Ames Research Center.

I thank an anonymous referee for providing very helpful suggestions and constructive criticisms. 
I acknowledge the support of the Basic Science Research Program through the National Research Foundation (NRF) of Korea 
funded by the Ministry of Education (NO. 2016R1D1A1A09918491).  I was also partially supported by a research grant from the 
NRF of Korea to the Center for Galaxy Evolution Research (No.2017R1A5A1070354).

\clearpage
%%%%%%%%%%%%%%%%%%%%%%%%%%%%%%%%%%%%%%%%%%%%%%%%%%
%%%%%%%%%%%%%%%%%%%%%%%%%%%%%%%%%%%%%%%%%%%%%%%%%%
\clearpage
%%%%%%%%%%%%%%%%%%%%%%%%%%%%%%%%%%%%%%%%%%%%%%%%%%
%%%%%%%%%%%%%%%%%%%%%%%%%%%%%%%%%%%%%%%%%%%%%%%%%%
\clearpage
%%%%%%%%%%%%%%%%%%%%%%%%%%%%%%%%%%%%%%%%%%%%%%%%%%
%%%%%%%%%%%%%%%%%%%%%%%%%%%%%%%%%%%%%%%%%%%%%%%%%%
\clearpage
%%%%%%%%%%%%%%%%%%%%%%%%%%%%%%%%%%%%%%%%%%%%%%%%%%
%%%%%%%%%%%%%%%%%%%%%%%%%%%%%%%%%%%%%%%%%%%%%%%%%%
\clearpage
%%%%%%%%%%%%%%%%%%%%%%%%%%%%%%%%%%%%%%%%%%%%%%%%%%
%%%%%%%%%%%%%%%%%%%%%%%%%%%%%%%%%%%%%%%%%%%%%%%%%%
\clearpage
%%%%%%%%%%%%%%%%%%%%%%%%%%%%%%%%%%%%%%%%%%%%%%%%%%
%%%%%%%%%%%%%%%%%%%%%%%%%%%%%%%%%%%%%%%%%%%%%%%%%%
\clearpage
%%%%%%%%%%%%%%%%%%%%%%%%%%%%%%%%%%%%%%%%%%%%%%%%%%
%%%%%%%%%%%%%%%%%%%%%%%%%%%%%%%%%%%%%%%%%%%%%%%%%%
\clearpage
%%%%%%%%%%%%%%%%%%%%%%%%%%%%%%%%%%%%%%%%%%%%%%%%%%
%%%%%%%%%%%%%%%%%%%%%%%%%%%%%%%%%%%%%%%%%%%%%%%%%%
\clearpage
%%%%%%%%%%%%%%%%%%%%%%%%%%%%%%%%%%%%%%%%%%%%%%%%%%
%%%%%%%%%%%%%%%%%%%%%%%%%%%%%%%%%%%%%%%%%%%%%%%%%%
\clearpage
%%%%%%%%%%%%%%%%%%%%%%%%%%%%%%%%%%%%%%%%%%%%%%%%%%
%%%%%%%%%%%%%%%%%%%%%%%%%%%%%%%%%%%%%%%%%%%%%%%%%%
\clearpage
%%%%%%%%%%%%%%%%%%%%%%%%%%%%%%%%%%%%%%%%%%%%%%%%%%
%%%%%%%%%%%%%%%%%%%%%%%%%%%%%%%%%%%%%%%%%%%%%%%%%%
\clearpage
%%%%%%%%%%%%%%%%%%%%%%%%%%%%%%%%%%%%%%%%%%%%%%%%%%
%%%%%%%%%%%%%%%%%%%%%%%%%%%%%%%%%%%%%%%%%%%%%%%%%%
\clearpage
%%%%%%%%%%%%%%%%%%%%%%%%%%%%%%%%%%%%%%%%%%%%%%%%%%
%%%%%%%%%%%%%%%%%%%%%%%%%%%%%%%%%%%%%%%%%%%%%%%%%%
\clearpage
%%%%%%%%%%%%%%%%%%%%%%%%%%%%%%%%%%%%%%%%%%%%%%%%%%
%%%%%%%%%%%%%%%%%%%%%%%%%%%%%%%%%%%%%%%%%%%%%%%%%%
\clearpage
%%%%%%%%%%%%%%%%%%%%%%%%%%%%%%%%%%%%%%%%%%%%%%%%%%
%%%%%%%%%%%%%%%%%%%%%%%%%%%%%%%%%%%%%%%%%%%%%%%%%%
\clearpage
%%%%%%%%%%%%%%%%%%%%%%%%%%%%%%%%%%%%%%%%%%%%%%%%%%
%%%%%%%%%%%%%%%%%%%%%%%%%%%%%%%%%%%%%%%%%%%%%%%%%%
\clearpage
%%%%%%%%%%%%%%%%%%%%%%%%%%%%%%%%%%%%%%%%%%%%%%%%%%
%%%%%%%%%%%%%%%%%%%%%%%%%%%%%%%%%%%%%%%%%%%%%%%%%%
\clearpage
%%%%%%%%%%%%%%%%%%%%%%%%%%%%%%%%%%%%%%%%%%%%%%%%%%
%%%%%%%%%%%%%%%%%%%%%%%%%%%%%%%%%%%%%%%%%%%%%%%%%%
\clearpage
%%%%%%%%%%%%%%%%%%%%%%%%%%%%%%%%%%%%%%%%%%%%%%%%%%
%%%%%%%%%%%%%%%%%%%%%%%%%%%%%%%%%%%%%%%%%%%%%%%%%%
\clearpage
%%%%%%%%%%%%%%%%%%%%%%%%%%%%%%%%%%%%%%%%%%%%%%%%%%
%%%%%%%%%%%%%%%%%%%%%%%%%%%%%%%%%%%%%%%%%%%%%%%%%%
\clearpage
%%%%%%%%%%%%%%%%%%%%%%%%%%%%%%%%%%%%%%%%%%%%%%%%%%
%%%%%%%%%%%%%%%%%%%%%%%%%%%%%%%%%%%%%%%%%%%%%%%%%%
\clearpage
%%%%%%%%%%%%%%%%%%%%%%%%%%%%%%%%%%%%%%%%%%%%%%%%%%
%%%%%%%%%%%%%%%%%%%%%%%%%%%%%%%%%%%%%%%%%%%%%%%%%%
\clearpage
%%%%%%%%%%%%%%%%%%%%%%%%%%%%%%%%%%%%%%%%%%%%%%%%%%
%%%%%%%%%%%%%%%%%%%%%%%%%%%%%%%%%%%%%%%%%%%%%%%%%%
\clearpage
%%%%%%%%%%%%%%%%%%%%%%%%%%%%%%%%%%%%%%%%%%%%%%%%%%
%%%%%%%%%%%%%%%%%%%%%%%%%%%%%%%%%%%%%%%%%%%%%%%%%%
\clearpage
%%%%%%%%%%%%%%%%%%%%%%%%%%%%%%%%%%%%%%%%%%%%%%%%%%
%%%%%%%%%%%%%%%%%%%%%%%%%%%%%%%%%%%%%%%%%%%%%%%%%%
\clearpage
%%%%%%%%%%%%%%%%%%%%%%%%%%%%%%%%%%%%%%%%%%%%%%%%%%
%%%%%%%%%%%%%%%%%%%%%%%%%%%%%%%%%%%%%%%%%%%%%%%%%%
\clearpage
%%%%%%%%%%%%%%%%%%%%%%%%%%%%%%%%%%%%%%%%%%%%%%%%%%
%%%%%%%%%%%%%%%%%%%%%%%%%%%%%%%%%%%%%%%%%%%%%%%%%%
\clearpage
%%%%%%%%%%%%%%%%%%%%%%%%%%%%%%%%%%%%%%%%%%%%%%%%%%
%%%%%%%%%%%%%%%%%%%%%%%%%%%%%%%%%%%%%%%%%%%%%%%%%%
\end{document}